%% file: depth.tex
\pdfoutput=1

\documentclass{acmsiggraph}
\usepackage[margin=1.9in]{geometry} 
\usepackage{dsfont}
\usepackage{wrapfig}
\usepackage{upgreek}
\usepackage{amsmath,amsfonts,amssymb,amsthm}

\usepackage[bookmarks=false]{hyperref}

\usepackage{ifpdf}
\DeclareGraphicsExtensions{.png, .jpeg}

\usepackage{color}
\definecolor{lightgray}{rgb}{0.64, 0.64, 0.64}

\usepackage[absolute]{textpos}

\pdfauthor{none}
\title{Skeletal Representations and Applications\footnote{Technical Report, School of Computing Science, Simon Fraser University, SFU-CMPT TR 2012-55-1}}
\author{Andrea Tagliasacchi}

\input{tweaks.tex}

\input{symbols.tex}

\begin{document}
\input{figures/teaser.tex}
\maketitle
\input{abstract.tex}


\newpage
\tableofcontents
\newpage
\input{intro.tex}
\input{medial.tex}
\input{skeleton.tex}
\input{conclusions.tex}

\bibliographystyle{acmsiggraph}
\bibliography{depth}

\input{appendix.tex}
\end{document}

%% file: tweaks.tex
\newcommand{\Fig}[1]{Fig.~\ref{fig:#1}}
\newcommand{\Figure}[1]{Figure~\ref{fig:#1}}

\newcommand{\Section}[1]{Section~\ref{sec:#1}}

\newcommand{\Definition}[1]{Definition~\ref{def:#1}}

\newcommand{\QUOTE}[1]{``#1''}
\newcommand{\textquote}[1]{``THIS IS DEPRECATED''}

\newcommand{\seeappendix}{$^\dagger$}

\newcommand{\ignore}[1]{}

\theoremstyle{definition}
\newtheorem{defn}{Definition}[section]
\newenvironment{definition}[1][Definition]{\vspace{-.15in}\begin{quote}\begin{defn}} {\end{defn}\end{quote}}

%% file: symbols.tex
\newcommand{\MAT}{\textbf{MAT}}
\newcommand{\M}{\mathcal{M}}

\newcommand{\mpoints}{\mathcal{M}}

\newcommand{\mradii}{\mathcal{R}}
\newcommand{\ball}{\mathcal{B}}

\newcommand{\V}{\mathcal{V}}

\newcommand{\boundary}{\mathcal{S}}
\newcommand{\surface}{\boundary}
\newcommand{\curve}{\boundary}

\newcommand{\object}{\mathcal{O}}

\newcommand{\R}{\mathds{R}}

\newcommand{\surfacescalar}{f}

%% file: figures/teaser.tex
\teaser{
\begin{minipage}[t]{\linewidth}
\centering
\includegraphics[width=\linewidth]{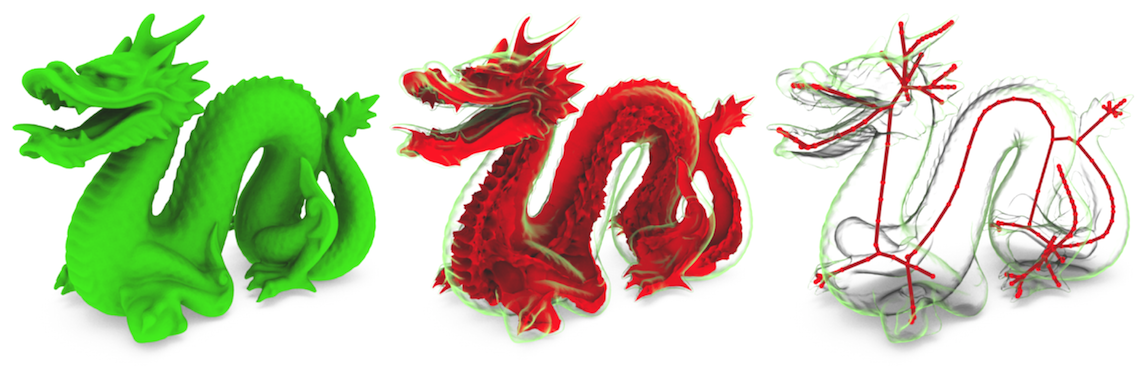}
\end{minipage}
\caption{Alternative representations of a solid shape: \textbf{(a)} a surface representations, \textbf{(b)} a medial skeleton representation, \textbf{(c)} a curve skeleton representation.}
\label{fig:teaser}}

%% file: abstract.tex
\begin{abstract}
When representing a solid object there are alternatives to the use of traditional explicit (surface meshes) or implicit (zero crossing of implicit functions) methods. Skeletal representations encode shape information in a mixed fashion: they are composed of a set of explicit primitives, yet they are able to efficiently encode the shape's volume as well as its topology. I will discuss, in two dimensions, how symmetry can be used to reduce the dimensionality of the data (from a 2D solid to a 1D curve), and how this relates to the classical definition of skeletons by Medial Axis Transform. While the medial axis of a 2D shape is composed of a set of curves, in 3D it results in a set of sheets connected in a complex fashion. Because of this complexity, medial skeletons are difficult to use in practical applications. Curve skeletons address this problem by strictly requiring their geometry to be one dimensional, resulting in an intuitive yet powerful shape representation. In this report I will define both medial and curve skeletons and discuss their mutual relationship. I will also present several algorithms for their computation and a variety of scenarios where skeletons are employed, with a special focus on geometry processing and shape analysis.
\end{abstract}

%% file: intro.tex

\section{Introduction}
\label{sec:intro}

Three-dimensional models of solid objects are commonly used in a large variety of disciplines: computer graphics, medical imaging, computer aided design, visualization and digital inspection, etc. The information contained in these models typically encodes whether a particular region of space is inside or outside the object (i.e. implicit/volumetric representations) or the interface between these two regions (i.e. explicit/surface representations). While they are invaluable in capturing details of the shape at an arbitrarily fine scale, more compact and expressive representations are often needed.

The advantages of compact representations are many. From a computational standpoint, we would like to represent the core information of our data in the most space-efficient way possible. This is beneficial as it allows algorithms that require exhaustive analysis of the input to work on much smaller datasets. The performance and quality of their results will depend on the conciseness and expressiveness of their description. On the other hand, many algorithms require some form of user intervention, like in editing or visualization of complex data. By providing the user with a compact and meaningful abstraction of the object, the complexity of this interaction can then be greatly reduced. 

\input{figures/symmetry.tex}

In order to be able to generate a compact representations we can observe that most objects present a certain degree of redundancy in the information they convey. Indeed, in a geometric setting, this redundancy can be measured by the \textit{symmetry} of the object. Intuitively, by removing the redundant information from an object that can be inferred through symmetry, we should be able to obtain more compact representations.

In an object, symmetry presents itself at different levels and by exploiting it we can create different levels of shape abstraction. For instance, consider the symmetries of the 2D solid object provided in \Fig{symmetry}. For simple shapes like $\object_a$, a single symmetry axis captures the global structure of the shape. However, such a global relationship is not always suitable. For example, by performing a simple quasi-rigid articulation of $\object_a$ we can obtain  $\object_b$, where a global symmetry axis is now insufficient. To address this problem we could use multiple, more localized, axes of symmetry. However, when the shape undergoes even more complex types of articulation, like $\object_c$, even the use of a limited set of axes of symmetry becomes too restrictive. The solution to this problem can be found by considering symmetry at its finest level. Each pair of points in the shape is linked by a \emph{infinitesimal symmetry relationship} and their centres of symmetry can be linked together to form a curvilinear axis commonly referred to as the \emph{skeleton} of the shape.

Historically, the extraction of skeletal representations can be dated back to Blum's work on the \emph{\textbf{M}edial \textbf{A}xis \textbf{T}ransform} \cite{blum_mit67}, which tracks the loci of maximally inscribed circles in two dimensional shapes. Although the MAT of a shape is unique, many alternative definitions exist, which I will discuss in \Section{medial:definitions}. Each of these definitions highlight different properties (\Section{medial:properties}), which in turn can be exploited for its computation (\Section{medial:computation}). Of all these alternatives the most noteworthy in the context of this discussion is the one provided by \cite{blum_jtb73}, which re-formulates the medial skeleton as the \emph{Symmetry Axis Transform} of the shape: an attempt to identify infinitesimal (i.e. local) reflectional symmetries.

In two dimensions, we can use the infinitesimal reflectional symmetries encoded by the MAT to create curvilinear skeletons. This simple network of curves provide us with the desired concise shape representation. Unfortunately, when we extend our attention to the three dimensional objects it retains its curvilinear characteristics only when we restrict our consideration to networks of tubular structures. For more general shapes, the medial skeleton, while still computable, is composed of sheets (two dimensional manifolds) stitched together in a complex fashion \cite{giblin_pami04}. Even though medial representations of 3D objects are still useful in many applications (\Section{medial:applications}), there are situations in which a simpler curvilinear representation is beneficial. These curvilinear representations of 3D shapes are commonly referred to as \emph{curve skeletons} and will be the subject of \Section{skeletons}.

Apart from medial skeletons, where unique and precise definitions (\Section{medial:definitions}) are available, there are multiple ways to define curve skeletons and their computation (\Section{skeletons}) is consequently application dependent. Typically we can think of the skeleton as a graph-like coding of the shape's essential structure (\Section{skeletons:properties}). This encoding attempts to represent shapes in ways that agree with human intuition: by representing connected components and the way they connect to form a whole. Thus, curve skeletons can be thought of as shape descriptors: a compact and expressive representation suitable for solving tasks which are computationally expensive if performed on large data sets.

%% file: figures/symmetry.tex
\begin{figure}[th]
\centering
\includegraphics[width=\linewidth]{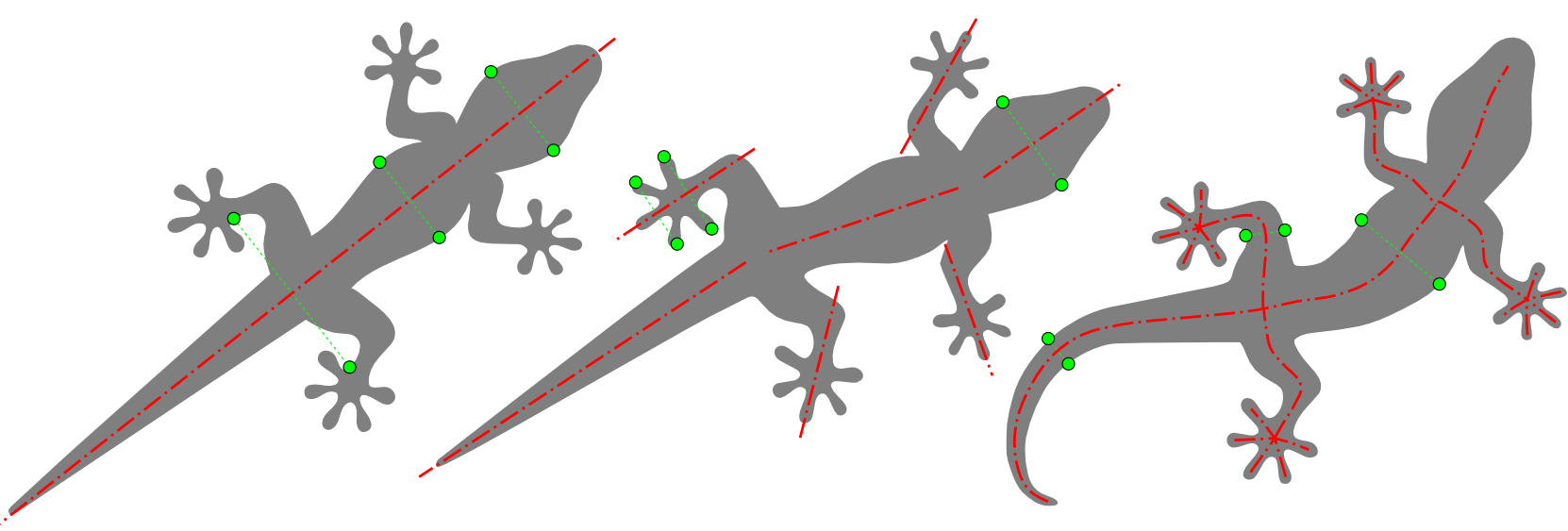} \\
\centering
\caption {Three planar solid objects $\object_a$, $\object_b$, $\object_c$ and their respective axes of symmetry. The planar symmetry relations (dotted green lines) between a few pairs of points (green dots) have been highlighted.}    
\label{fig:symmetry}
\end{figure}

%% file: medial.tex
\section{Medial skeletons}
\label{sec:medial}
\input{figures/medial.tex}

The concept of \emph{medial skeletons} originated in the context of two dimensional shape understanding. It was created as a way of reducing the large amount of information carried by a shape down to a \QUOTE{skeleton} of crucial information that can be more readily assimilated. In \Figure{medial} an example of medial skeleton for a planar shape is shown. Note how the medial skeleton is able to effectively capture the parts that compose the shape and the way in which they are interconnected to form the whole. 

Medial skeletons can be defined in many ways, as discussed in \Section{medial:definitions}. As the medial skeleton of a shape is unique, these definitions are equivalent and reporting just one of them unequivocally defines medial skeletons. Nevertheless, these alternative formulations are very informative: they reveal essential shape properties which I will further discuss in \Section{medial:properties}. These properties will offer key insights that will help provide the foundations for the creation of algorithms for skeleton extraction (\Section{medial:computation}) and their practical application (\Section{medial:applications}).

%
%
\subsection{Defining medial skeletons}
\label{sec:medial:definitions}
\input{figures/medial_definitions.tex}
In this section we will formally define medial skeletons of a shape. In this context a shape is nothing but a \emph{solid object} $\object \subset \R^n$  with \emph{boundary} $\boundary=\partial \object$. Note that while I will define them and discuss their properties predominantly in two dimensions, these, unless otherwise noted, generalize to three dimensions. In $\R^2$ we will refer to the $\boundary$ as the \QUOTE{contour} of the shape while in $\R^3$ it will be simply referred to as the \QUOTE{surface} of the object. Also note that while in general the medial skeletons can be defined both for $\object$ and its complement space $\R^n \setminus \object$, our focus will be on the skeletons of $\object$: the internal medial skeleton.
  
\paragraph{Maximally inscribed balls} 
The original idea of medial skeletons was introduced in the two dimensional domain by \cite{blum_mit67}. Blum proposed that the skeleton of an object $\object$ can be extracted by fitting maximally inscribed balls\seeappendix in $\object$ and noting the locus of their centres (see \Fig{medial:definitions}-a). By augmenting the location of these centres $\mpoints$ with the radius of the corresponding spheres $\mradii$ we obtain the \emph{Medial Axis Transform} $\left[ \mpoints, \mradii \right] = \MAT (\object)$.

\begin{definition}
    The Medial Axis Transform of $\object$ is the set of centres $\mpoints$ and radii $\mradii$ of all the maximal inscribed balls in $\object$.
    \label{def:medial:inscribed}
\end{definition}

Note that since we only consider balls which are \emph{maximal}, many inscribed balls and their associated loci will simply be discarded. The representation that we will obtain will consequently be sparse: the medial axis of $\object \subset \R^n$ has \emph{at most} dimensionality $n-1$. In addition, by superimposing medial balls of radius $\mradii$ at locations $\mpoints$ we can recreate the original shape $\object$. This allows us to understand how the MAT is indeed a \emph{transform} as it admits invertibility. 

While Blum's original definition provides us with much insight on its capabilities, it is hard to think in terms of maximally inscribed balls. For this reason a more commonly known definition is the one based on the \emph{grassfire analogy}.

\paragraph{Grassfire analogy}
Imagine $\object \subset \R^2 $ as if it were a patch of grass whose whole boundary $\surface$ is set on fire at $t=0$. The fire will propagate isotropically from the boundary towards the interior of the shape with uniform speed along the normals $\mathbf{n}$. At certain locations, the fire fronts coming from different parts of the shape will meet and quench, thus defining a shock graph \cite{kimia_ijcv95}:

\begin{definition} 
    The Medial Axis Transform of $\object$ with boundary $\surface$ is given by the shock graph of the motion $\dot{S}=-\mathbf{n}$  and the time $t$ at which each shock is formed.
    \label{def:medial:grassfire}
\end{definition}

This definition not only stands at the core of many medial axis computation algorithms, which we will analyze in \Section{medial:computation}, but it also allows us to intuitively understand why these skeletons are called \emph{medial}. As two different sides of the boundary move at the same speed, their quenching location is located deep in the object, along its centreline. 

\paragraph{Maxwell set}
As the grassfire propagates isotropically, quench points will always be equidistant from the boundary. Consequently medial points are always associated with at least two (Euclidean) closest points on $\surface$. This very property lies at the core of the \emph{Maxwell set} definition of Medial Axis Transform \cite{mather_spma83}:

\begin{definition}
    The Medial Axis Transform is the set of locations $\mpoints$ internal to the object with more than one corresponding closest boundary point and their distance $\mradii$ from the boundary $\surface$.
    \label{def:medial:maxwell}
\end{definition}

This definition has a key importance in the geometric setting. In \Section{medial:computation:voronoi} we will discuss how this lies at the basis of medial extraction algorithms based on the Voronoi diagram.

\paragraph{Symmetry axis}
Let us momentarily restrict our attention to objects with $C^1$ continuous boundaries. If we take a points $x \in \R^2$, then the closest point on the boundary $s^* \in \surface$ will define a ball $\ball_{x,r}$ centred at $x$ of radius $r=\| x-s^*(x) \|$. This ball can be proved to be tangent to $S$ at $s^*$ \cite{amenta_sig98}. If we restrict our attentions to medial points, which have at least two closest boundary points, then the associated medial balls will be bi-tangent to $S$:

\begin{definition}
    The Medial Axis Transform of $\object$ is the set of centres $\mpoints$ and radii $\mradii$ of all the inscribed balls in $\object$ which are bi-tangent to its boundary $\surface$.
    \label{def:medial:symmetry}
\end{definition}

Interestingly, the locus of bi-tangent balls was introduced by \cite{giblin_amm85} under the name of \emph{symmetry set}. Consequently, the medial axis can be seen as a subset of the symmetry set, where only inscribed balls are to be considered. This simple observation underlines how skeletons can be thought of as a representation of symmetry axis as discussed in \Section{intro}. Furthermore, note that while this equivalency required smooth boundaries, it extends to non-smooth ones by considering bi-tangency in the general sense.

%
%
\subsection{Medial axis properties}
\label{sec:medial:properties}

The definitions from \Section{medial:definitions} gave us much insight on what medial skeletons, or better, Medial Axis Transforms are. First, we understood why it's \emph{medial}, as it is located in the middle of the shape. Then we understood why it's an \emph{axis}, by observing its interpretation as an infinitesimal symmetry axis and its correlation to the symmetry set. But most importantly we understood it is a \emph{transform} as it is unique and fully invertible. Consequently, the Medial Axis Transform can be thought of as a dual representation of shape, which exhibits particular properties of the shape. In this section we will focus on three core properties: 1) its ability to compactly represent topology, 2) the association between medial branch and shape part and 3) the ability to encode local shape volume.

In two dimensions, skeletons have received much attention as they are very effective at capturing the \emph{topological properties} of the shape. Indeed, it can be shown how a shape and its medial axis are homotopy equivalent \cite{lieuter_sma03}. Although both volumetric and boundary representations have this ability, the skeleton is able to represent topology more compactly by means of a lower dimensional construct. Remember that for a $\R^n$ solid, the medial skeletons will have at most dimensionality $n-1$. Consequently, for two dimensional objects this will result in one-dimensional (curvilinear) medial skeletons. In a similar fashion, a three dimensional tubular shape can be represented by a medial skeleton as a collection of centrelines that split and join to encode the object topology. However, when we move our attention to 3D solids, medial skeletons will not be curvilinear, as they can contain up to $(3-1)$D construct: a collection of two-manifolds, curves and points in space (see \Figure{teaser}-b). For this reason the representation of topology in three dimensional data is more commonly performed by curve skeletons (see \Figure{teaser}-c) discussed in \Section{skeletons}.

\paragraph{Medial branches and the \QUOTE{curse of instability}}
\input{figures/medial_twodstability.tex} 
By observing \Figure{medial} we can see how the medial skeleton does not exclusively capture the topology of the shape. Indeed, there would be no topological difference in between the tuning fork and any other planar shape homotopic to a disk. In fact, the medial is composed of a set of branches which are descriptive of the shape structure -- it decomposes a shape in parts and they way in which they are interconnected (\Figure{medial}-c). In order to better understand the limitations of medial skeletons, it is important to see what causes the creation of skeletal branches. To analyze this behaviour we can restrict our attention to planar polygons, whose medial axis transform simply consists of straight lines and parabolic arcs. In this simplified scenario, illustrated in \Figure{medial:twodstability}-a, we have a straightforward relationship: a medial branch can be found for every convex vertex of the boundary. This is indeed a cause of concern, since as illustrated in \Figure{medial:twodstability}-(b,c) polygonal objects with very similar geometry can result in very different medial skeletons. Ideally we would desire to have a branch for any \QUOTE{important} geometric part/feature of the shape. Here, the concept of importance is to be intended in terms of relative geometric scale w.r.t. other shape features. Note however that the classical medial skeleton is scale insensitive: the addition of an arbitrarily small geometric feature results in large, yet local, change on the medial axis. This property (or lack thereof) is known in the literature as the \emph{stability} of the medial axis~\cite{attali_mfsvcgdv09} and will be the subject of \Section{medial:computation:stability}.

\paragraph{Correspondence and local thickness}
By observing \Figure{medial:definitions}-c we can perceive the existence of a very natural link between points on medial axis surfaces and points on the object boundary surfaces. \Definition{medial:maxwell} allowed us to understand the first of these links: for each point on the medial axis there is \emph{one or more} corresponding points on the boundary of the object. It is interesting to note that for at least $C^1$ smooth boundaries the opposite relation is quite different as for each point on the object boundary there is \emph{exactly one} corresponding point on the medial surfaces. This observation is extremely important because it allows us to attach to each of the boundary points the size of the corresponding medial ball $\ball$. The injectivity of the correspondence allows us to transport the scalar function $\mradii$ defined on $\mpoints$ onto the shape surface $\boundary$. This scalar function defined on the surface is extremely important as it provides a formal way to define local shape thickness whose applications will be explored in \Section{medial:applications}.

%
%
\subsection{Medial axis computation}
\label{sec:medial:computation}
A large body of work exists on medial axis computation. In this section I will briefly describe the core idea behind these solutions. In geometry processing objects are most commonly represented by their boundary. For this reason I will introduce and describe methods that are progressively more suitable for these representations. 


\subsubsection{Exact medial computation}
Regardless of the representation, if we are given an \emph{exact} description of the shape then it should also be possible to compute an exact medial axis representation. Typically these exact computations follow the intuition behind the Maxwell set definition. For example if the boundary of an object is represented as polygons, then we can extract the medial axis by computing their bisectors as shown by \cite{lee_pami82,held_cgta01} in two dimensions and \cite{culver_cagd04} in three dimensions. Note that the extraction of the exact medial axis is not restricted to piecewise linear representations. Recently the authors of \cite{tzoumas_ecg11} demonstrated how to compute the medial axis of rational curves in the plane, while \cite{musuvathy_cad11} addressed a similar problem in the three dimensional domain. 

The extraction of exact medial representations, albeit possible, suffers from two major shortcomings. First of all, none of the methods above is able to deal with large datasets, like the ones typically found in most practical applications. This complexity is due to the fact that to compute the exact medial axis we have to solve systems of high degree algebraic equations. But even more importantly, in most scenarios geometric shapes are not known exactly. Shapes are indeed represented by different kinds of approximation, thus exactly extracting the medial axis is often unnecessary. For these reasons we will now 
move our attention to algorithms for approximate medial axis extraction and discuss the quality of this approximation.


\subsubsection{Topological thinning}
\label{sec:medial:computation:thinning}
In the discussion of \Definition{medial:inscribed} I highlighted the fact that the medial axis is a compact structure, with a much smaller volume than the original object, and how this structure is necessarily contained within its boundaries. If we represent our shape as a set of voxels sampling its interior, then a simple idea to compute medial axis is to progressively reduce the volume from the outside-in until a \QUOTE{thin} axial representation of unitary thickness is obtained. This criterion lies at the base of topological thinning methods like the one in \cite{palagyi_gmip99}. 

An atomic reduction in volume can simply be achieved by removing a voxel from the set. This removal has to be done carefully so not to alter the overall topology of the solid. Therefore, algorithms in this class typically have to specify complex local measures of connectedness for both interior and exterior. Also, as we generally desire skeletons that are \emph{medial} within the shape, this removal needs to be performed isotropically: by simultaneously removing voxels from the whole surface. However, in highly discrete settings, perfect isotropy is hard to achieve. Consequently, most algorithms suffer of the drawback that the structure of the resulting set depends on the voxel processing order, often resulting in skeletons which are non smooth. In addition, in regions whose thickness is expressed by an even number of pixels/voxels, the set of centres of maximal balls is two voxels wide. In these regions further thinning is possible but this often results in an even rougher axis. 

For all these reasons, thinning methods are far from optimal for computing medial axis of detailed surface geometry as we would incur in two levels of approximation: in the surface to volume conversion as well as in the computation of the axis.

\subsubsection{Distance fields}
\label{sec:medial:computation:dfields}
Rather than working directly on a discretized representation of the surface it is often convenient to represent the shape by means of an Euclidean implicit function: a real valued function that stores the Euclidean distance from the surface to any point in space. As illustrated in \Fig{medial:definitions}-b the iso-lines of the distance field correspond to positions of the boundary undergoing normal motion. Exploiting this observation, distance field methods attempt to estimate the medial axis by analyzing the local properties of the distance transform. Of particular interest are its ridges: the regions where its derivatives are multi-valued, which correspond with medial shock graphs.

Once the distance transform is computed, the extraction voxel along its ridges can be performed in a number of ways. In \cite{sanniti_ivc96} the authors first detect critical points in the distance field and then connect them to produce medial skeletons. Alternative methods identify medial voxels by thresholding some scalar function computed from the distance field. These function generally describe the likelihood of the voxel to be located on the ridge of the distance transform, like the \emph{average inward flux} in \cite{siddiqi_ijcv02} 
or the \emph{object angle extension} of \cite{shah_cviu05}. In many scenarios a simple thresholding is not enough as it would corrupt topology. For this reason,  distance field methods are often coupled with the topology preserving thinning procedures discussed in \Section{medial:computation:thinning}.

Even though implicit functions are most commonly stored on uniform discrete grids, they offer a much better shape approximation quality than simple voxelization, as typically the distance to the surface can be interpolated within each voxel. Furthermore, while the complexity for the computation of the Euclidean distance transform by fast-marching \cite{sethian_nas96} is typically $O(n\log n)$, it is often acceptable to use other computationally cheaper types of distances. For example, in \cite{sanniti_ivc96} by using \emph{city-block distance}, we are able to reduce the complexity of this pre-processing operation down to $O(n)$.

While the domain is represented more accurately, some of the limitations of thinning methods apply to distance transform methods as well. The medial axis is represented by a set of discrete voxels, and it is still possible to find regions of non-unitary thickness, making it a less than optimal choice for medial axis computation of surface representations. 

\subsubsection{Voronoi methods}
\label{sec:medial:computation:voronoi}
\input{figures/medial_voronoi.tex}
The Maxwell set interpretation of the medial axis addressed in \Definition{medial:maxwell} provides us with what we need to formulate the medial axis computation as a \emph{Voronoi diagram} problem. This connection is immediately understood in two dimensions, as a Voronoi edge (vertex) is the locus of points that has two (three) closest neighbors. As the Voronoi diagram input is a point set, the boundary $\surface$ of the object $\object$ typically needs to be sampled. Once a sampling is obtained, we can simply compute its Voronoi diagram $\V$. As illustrated in \Figure{voronoimedial}, the medial axis can then be approximated by an appropriate subset of $\V$. In this section we will discuss the importance of the sampling and how to choose a proper subset 
to obtain good medial approximations.

\input{figures/medial_voronoi_twod.tex}

It was Blum in his original work \cite{blum_mit67} that observed how, for a dense enough sampling of a smooth planar curve, the corresponding Voronoi diagram appeared to approximate $\M$. As illustrated in \Figure{medial:voronoi:twod}, the denser the sampling of the curve, the better the medial approximation achievable. However, it was not until much later that \cite{brandt_cvgip92} proved that, given a dense enough uniform $\delta$-sampling\seeappendix, the subset of Voronoi fully contained within $\curve$ correctly approximates the medial axis with a convergence guarantee. This result was then extended by \cite{attali_cviu97} where the requirement of an oracle knowledge of $\surface$ was removed, and a proven topologically correct medial approximation was shown possible from a sampling of the surface only. Further extension was provided by \cite{amenta_gmip98}, which suggested how uniformly sampling the boundary is often unnecessary and an adaptive $\epsilon$-sampling\seeappendix can be used.

\input{figures/medial_voronoi_threed.tex}
Unfortunately, when we expand our interest to three dimensional objects, the results of \cite{attali_cviu97} or \cite{amenta_gmip98} do not directly hold. The problem is that, even for arbitrarily fine samplings, the Delaunay triangulation dual to the Voronoi diagram often presents sliver tetrahedra. These sliver tetrahedra correspond to Voronoi vertices, which neither fall in proximity of the medial axis nor are related to any prominent feature of the surface. To address this issue, \cite{amenta_scg98} approximated the medial axis by a more restrictive subset of the Voronoi diagram: by only considering the \emph{Voronoi poles}\seeappendix. The validity of Voronoi poles for medial approximation was formally verified by \cite{amenta_cgta00}, where the authors showed that for an $\epsilon$-sampled $C^1$ manifold, the poles approach the medial axis of the shape as $\epsilon$ vanishes. The validity of Voronoi poles is intuitively illustrated in \Figure{medial:voronoi:threed} where we visually compare their approximation power to the one of internal Voronoi vertices.


\subsubsection{Stable medial axis}
\label{sec:medial:computation:stability}
\input{figures/medial_filtering.tex}

When using medial representations for shape processing, we would like for similar shapes to have similar medial representations. However, while the Medial Axis Transform provides a perfectly invertible (i.e. dual) representation, a major shortcoming is that, as discussed in \Section{medial:properties}, it is not \emph{stable} with regards to small geometric variations. This issue is particularly important when we consider samples that are noisy, as even slight amounts of noise can produce long spurious branches in the MAT. This noise can be caused by a measurement process, when samples have been acquired by scanning the surface of real a world object, or by the fact that we are sampling a discretization of the shape. Here, I will review a few selected alternatives for noise management while referring the reader to \cite{attali_mfsvcgdv09} for a more extensive coverage of the topic.

Typically, a simpler medial axis can be obtained by performing a smoothing operation, however, controlling the right amount of smoothing has been shown to be a very challenging problem~\cite{attali_mfsvcgdv09}. For this reason, a large majority of the literature addresses the problem by considering the stability of a medial point either in terms of its local or global properties.

\paragraph{Local filtering criteria}
One common way of creating medial skeletons is by pruning its branches according to some \emph{local} criteria. A traditional choice lies in using some local measurement of a medial sample like the corresponding pair of points on the surface of \Figure{medial:definitions}-c. For example, \cite{chazal_sma04,chazal_gm05,chaussard_dcgi09} proposed the $\lambda$-medial axis, where medial samples whose corresponding surface samples have a \emph{circumradius} larger than $\lambda$ are discarded. For particular choices of the $\lambda$ parameter \cite{chazal_scg05}, the medial axis approximation is homotopic to the shape, and its approximation quality is provably convergent. Unfortunately, as it is illustrated in \Fig{medial:filtering}-b, this criterion doesn't allow the capture of details across different scales, as important features of the shape are removed before eliminating all of the noise. An alternative choice was presented by \cite{attali_icip96,amenta_sma01,dey_cad04,sud_spm05}, where the authors use the spoke aperture angle as a filtering criterion. However, while this approach is able to retain features at different scales, it might result in an axis whose topology is drastically different as we change the thresholding parameter as it is illustrated in \Figure{medial:filtering}-c. The common problem is that since these solutions are highly local it allows them to be efficient and simple to analyze but unable to capture the multi-scale structure of the shape.

\paragraph{Global filtering criterions}
Following a more global approach, we can compute the multi-scale importance of a medial branch based on volumetric approximations of the shape like in \cite{tam_vis03} or by analyzing its global connectivity to maintain topological equivalence to the shape \cite{sud_spm05}. Alternatively, we can address the problem in a statistical sense by analyzing the resilience of a medial branch in a collection of shapes like in the works of \cite{styner_ijcv03} and \cite{ward_pami09}. However, because of its simplicity, the most interesting global solution to the problem is the \emph{Scale Axis Transform} proposed in 2D by \cite{giesen_scg09} and successfully applied to 3D data by \cite{miklos_tog10}. In the scale axis, the global relationship between parts is obtained by scaling the medial balls of the shape by a factor $\delta$. Small medial balls, associated with small surface features and noisy medial branches, undergo a growth that is smaller than nearby large medial balls (i.e. larger nearby features). Following \Definition{medial:inscribed}, whenever one of these scaled balls is not maximally inscribed, it is removed from the set. A stable medial axis is then computed as the exact medial axis of the remaining set of de-scaled balls. As illustrated in \Figure{medial:filtering}-e, the quality of the resulting medial axis is far superior to the one of more local methods.

\subsubsection{Approximate medial measures}
While the methods discussed up to this point provably produce convergent approximations of the medial axis, there are situations in which it is possible to compute coarser estimates. For example, in situations in which only a rough approximation of the \emph{medial surfaces} is desired, the shape surface triangulation can be recycled like in the manifold medial approximation introduced by \cite{hisada_cgf02}. Alternatively, in situations in which only an estimate of \emph{local thickness} is required, the relatively expensive medial computation can be replaced by an (embarrassingly parallel) ray casting procedure \cite{shapira_vc08}. The \emph{distance field} methods of \Section{medial:computation:dfields} can exploit coarse-to-fine spatial discretizations, where efficient yet not-as-rigorous algorithms for thresholding can be used \cite{bloomenthal_sma99}. Differently from \Section{medial:computation:stability}, the generation of \emph{stable representations} can be aided by the use of templates offering pre-determined connectivity \cite{pizer_ijcv03} or by softening some of the strict requirements imposed by the medial axis transform \cite{pizer_dagstuhl11}.

%
\subsection{Medial axis applications}
\label{sec:medial:applications}
\input{figures/medial_applications_mix.tex}

The properties of the medial representation do not only lay the foundations for its computation but also give us the fundamental intuition to understand the many applications it has in practical scenarios. 

The \emph{volumetric} capabilities of the MAT has been exploited in a number of ways. In the medical domain, it has been used as a way of measuring local volumetric properties of the shape \cite{naf_cviu97}, as well as for segmentation by fitting volumetric templates and statistical analysis \cite{pizer_ijcv03}. Alternatively, in 3D the local volume can be clustered to segment the shape in patches of similar thickness as shown by \cite{shapira_vc08}, used as a prior for hole filling \cite{tagliasacchi_sgp11} or to facilitate shape modeling of organic shapes \cite{angelidis_sma02}. The \emph{medial} positioning of its curves can be exploited as an alternative to implicit functions to compute shape blends \cite{blanding_cg00}, as well as for navigation purposes \cite{paik_mp98}, while their interpretation as a set of \emph{inscribed spheres} can be used to generate a simplified model for collision detection \cite{hubbard_tog96}. In 3D, the boundaries of medial surfaces have been shown to be associated with important perceptual features \cite{hisada_cgf02}, which can be exploited to reduce the complexity of registration tasks \cite{cheng_pvt04}. As suggested by \cite{storti_sma97}, the generation of \emph{stable} medial axis can be exploited in the generation of shape simplification schemes from a volumetric point of view like in \cite{tam_vis03}, while the continuous connection existing between medial surfaces and shape surface \cite{shaham_sma04} describes the volume in a way that can be exploited for the generation of hexahedral meshes \cite{tchon_imr03}.

Perhaps the most important characteristic of the medial axis is that, in two dimensions, it is able to reduce the shape to a network of 1D curves. For 2D shapes, this \emph{graph-like} representation lies at the core of a large set of  algorithms that address the important problems of shape retrieval and correspondence \cite{siddiqi_ijcv99}. Unfortunately, these techniques do not directly apply to 3D shape because, as discussed in \Section{medial:properties}, their medial axis in the general setting is not composed of 1D curves. Thus, to be able to apply these methods with minimal modifications, we need to be able to generate \emph{curvilinear} representations for 3D shapes. These curvilinear representations are called \emph{curve skeletons} and their computation and applications will be the subject of \Section{skeletons}.    

We will conclude this section by describing two core applications of medial axis. These applications are possible thanks to the fact that the MAT offers a fully equivalent (i.e. dual) representation of the surface of an object. 

%
\subsubsection{Surface reconstruction}
\input{figures/medial_applications_surfrecon.tex}  
Given a sampling of the surface of a solid object, the problem addressed by surface reconstruction is the one of obtaining a parametrized (e.g. triangulated) representation of the surface.

In this context the medial axis was shown by \cite{amenta_scg98} to be essential in formalizing the minimal requirements of sample density. Intuitively, we would like a larger sampling density in regions of \emph{high curvature}, so as to be able to capture minute surface variations. At the same time, an increased number of samples is needed in \emph{narrow regions} to properly capture the shape's topology. To address both these requirements, \cite{amenta_scg98} introduced the $\epsilon$-sampling, where the necessary local sample density depends on the \emph{local feature size}: the distance to the closest point on the medial axis.

The contributions of medial representations do not end here. In particular, a way to approach the reconstruction problem is by computing the Delaunay triangulation and discarding a subset of its elements \cite{edelsbrunner_sig94}.  In \cite{amenta_sig98}, the authors extended this approach in an intuitive and powerful way: given an appropriate $\epsilon$-sample, they show that, by discarding elements whose circumsphere contains a portion of the medial axis, results in a topologically correct reconstruction of $\object$. This method and its essential relationship with the medial axis created the basis for much work in the field of combinatorial reconstruction \cite{cazals_ecg06}. 

\subsubsection{Shape manipulation}
\input{figures/medial_applications_shapemanip.tex}
Developing geometry-manipulation tools for end users is a core problem in geometry processing. The medial axis offers an alternative to the more commonly used surface based methods by exploiting the volumetric information of the shape captured in the axis.

In \cite{storti_sma97} the authors propose to use the medial domain as an alternative to surfaces for shape parameterization, thus creating the foundations for the widely used m-reps of \cite{pizer_tech00} and the shape modeling frameworks of \cite{angelidis_sma02}. Intuitively, by sparsely sampling the medial axis, the shape can be manipulated by modifying the position and radius of medial loci and then reconstructing/deforming the object surface. 

Following a similar paradigm, \cite{bloomenthal_sma99} investigated a way of connecting the use of geometric (medial) skeletons to traditional kinematic skeletons for shape animation. Rather than directly expressing surface vertex positions as a function of the kinematic skeleton, the author proposes to first animate the medial skeleton and then reconstruct the surface from its medial representation. Further connections between these two structures were also investigated by \cite{baran_sig07}, where instead of attempting to model surface animation by means of medial axis, the authors consider the medial domain an appropriate space for fitting a template of a kinematic skeleton. Traditionally, the transformation of animated surface vertices is obtained by a linear combination of transformations defined on kinematic skeletons. While these weights are typically set by artists by a time consuming trial and error process, \cite{bloomenthal_sca02} uses the medial axis and the radius function defined thereon to create a convolution scheme that is able to completely automatically create blending weights, resulting in natural-looking surface animations.

Rather than directly modifying the medial axis, or use a kinematic skeleton as a control structure, in \cite{yoshizawa_sma03,yoshizawa_eg07}, the authors investigated the use of freeform deformation of the medial surfaces for shape deformation. Their solution presents two strong advantages: first of all, thanks to the volumetric nature of medial representation, more natural large scale deformations can be achieved as local thickness is preserved; furthermore, self-intersections, which often arise when large deformations take place, can efficiently be corrected by exploiting the volumetric nature of medial representations.

%% file: figures/medial.tex
\begin{figure}[th]
\centering
\includegraphics[width=\linewidth]{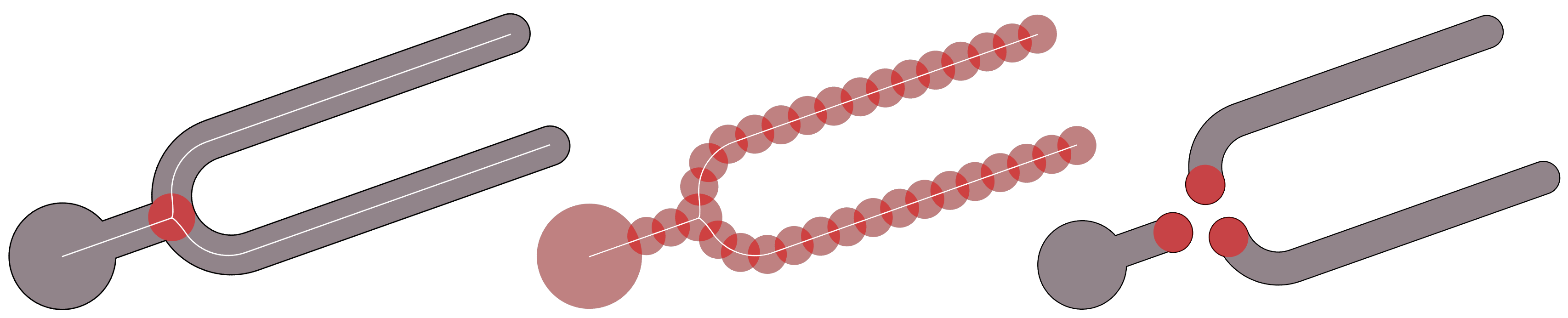} \\
\caption {Medial Axis Transform essentials: \textbf{(a)} a planar solid object (gray), its medial axis (white) and one of its maximal inscribed balls (red); \textbf{(b)} the superposition of a subset of medial balls and its ability to capture the original geometry; \textbf{(c)} the branch structure of the medial axis can be used to decompose the shape in meaningful parts.}  
\label{fig:medial}
\end{figure}

%% file: figures/medial_definitions.tex
\begin{figure}[tH!]
\centering           \includegraphics[width=\linewidth]{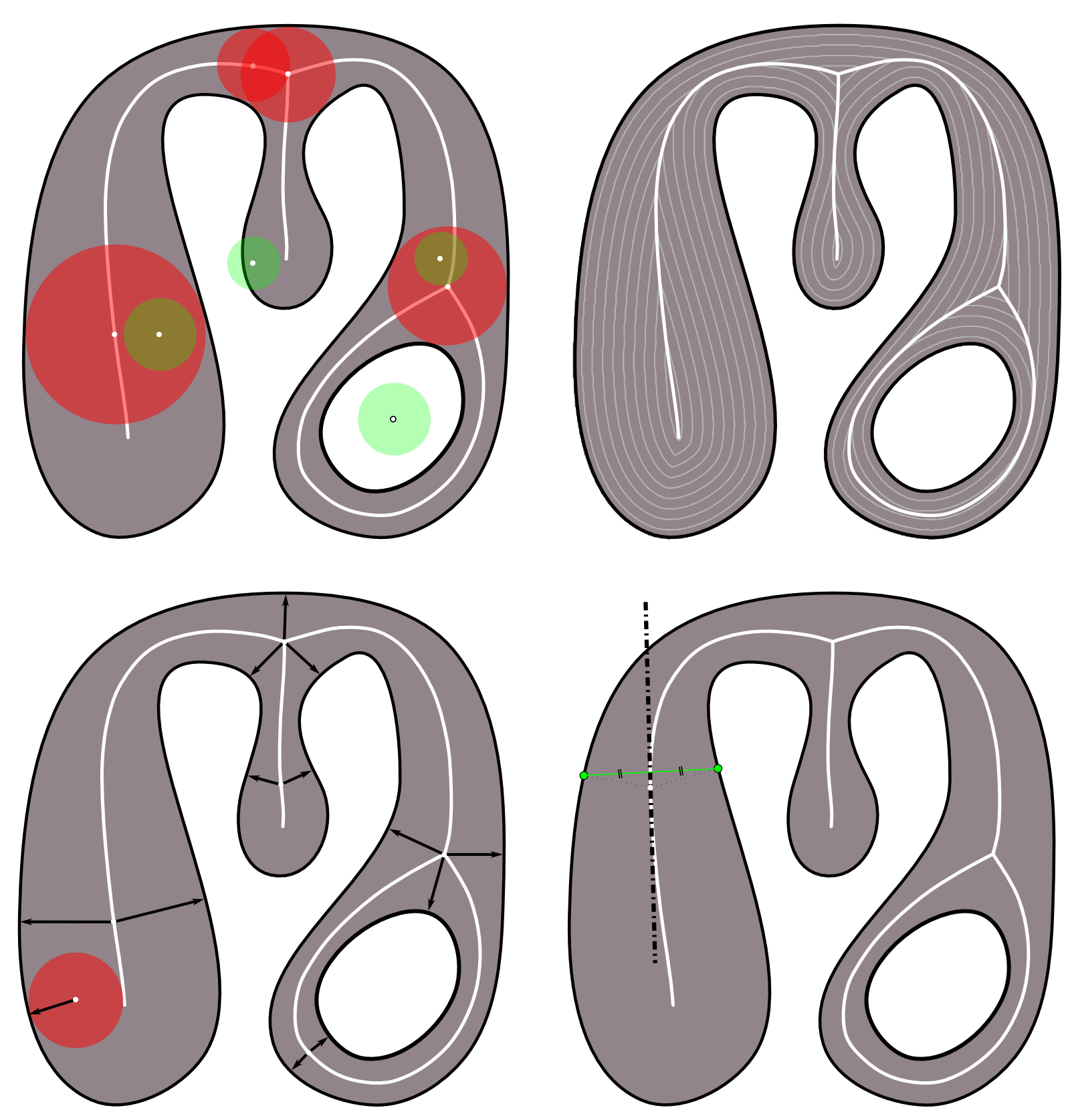}
\caption {The four possible definitions of medial axis, in lexicographical order: \textbf{(a)} only the centres of maximally inscribed balls, colored in red, are part of the medial axis; \textbf{(b)} the medial axis as the shock graph of grass-fire evolution, the location of the boundary at different times is illustrated by thin white lines; \textbf{(c)} only points on the medial axis have more than one corresponding point on the shape boundary; \textbf{(d)} the medial axis links boundary points with an infinitesimal reflectional symmetry relationship.}    
\label{fig:medial:definitions}
\end{figure} 

%% file: figures/medial_twodstability.tex
\begin{figure}[th]
\centering
\includegraphics[width=\linewidth]{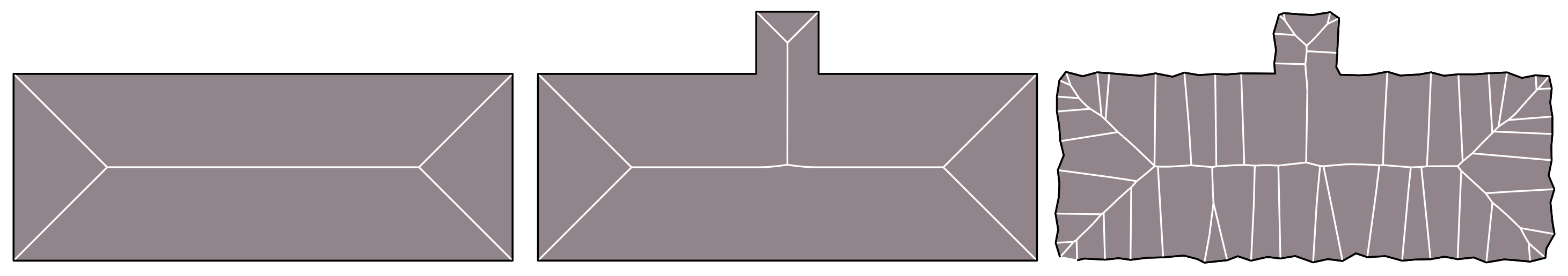}
\caption {\textbf{(a)} Medial axis of a simple rectangular polygon. \textbf{(b)} Inserting a new feature creates localized changes in the skeletal structure. \textbf{(c)} The classical medial axis is unstable w.r.t. small scale features (i.e. noise)}.
\label{fig:medial:twodstability}
\end{figure}

%% file: figures/medial_voronoi.tex
\begin{figure}[th]
\centering
\begin{minipage}[t]{0.49\linewidth}
\centering    	\includegraphics[width=\linewidth]{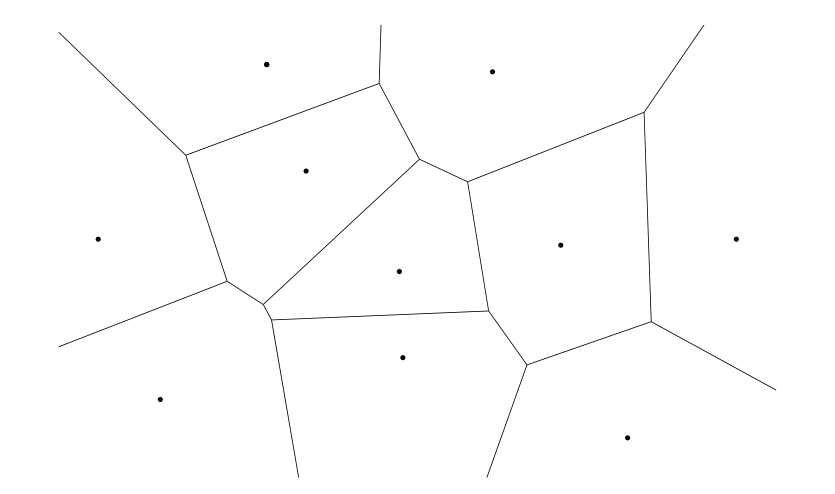} \\
\end{minipage}
\centering
\begin{minipage}[t]{0.49\linewidth}
\centering    	\includegraphics[width=\linewidth]{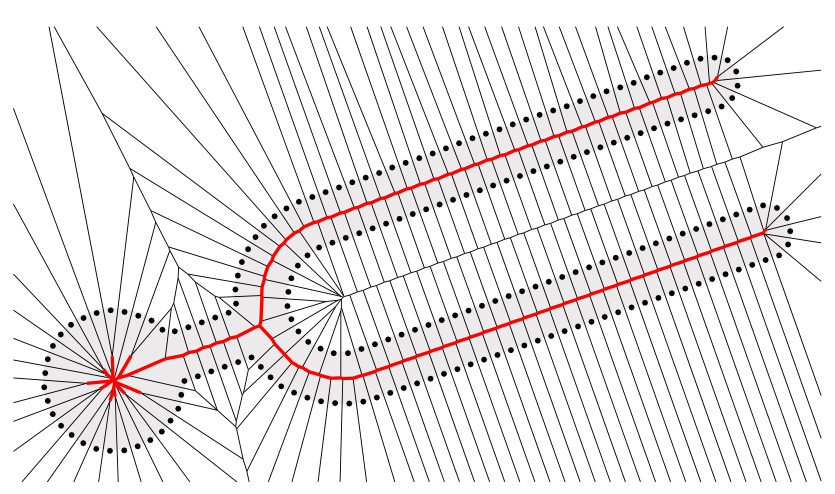} \\
\end{minipage}

\caption {\textbf{(a)} The Voronoi diagram of six points in the plane. \textbf{(b)} The Voronoi diagram of points sampled from the boundary of a shape. Note how Voronoi vertices and edges fully contained in the shape boundary form an approximation of the medial axis.}
\label{fig:voronoimedial}
\end{figure}

%% file: figures/medial_voronoi_twod.tex
\begin{figure}[th]
\centering
\includegraphics[width=\linewidth]{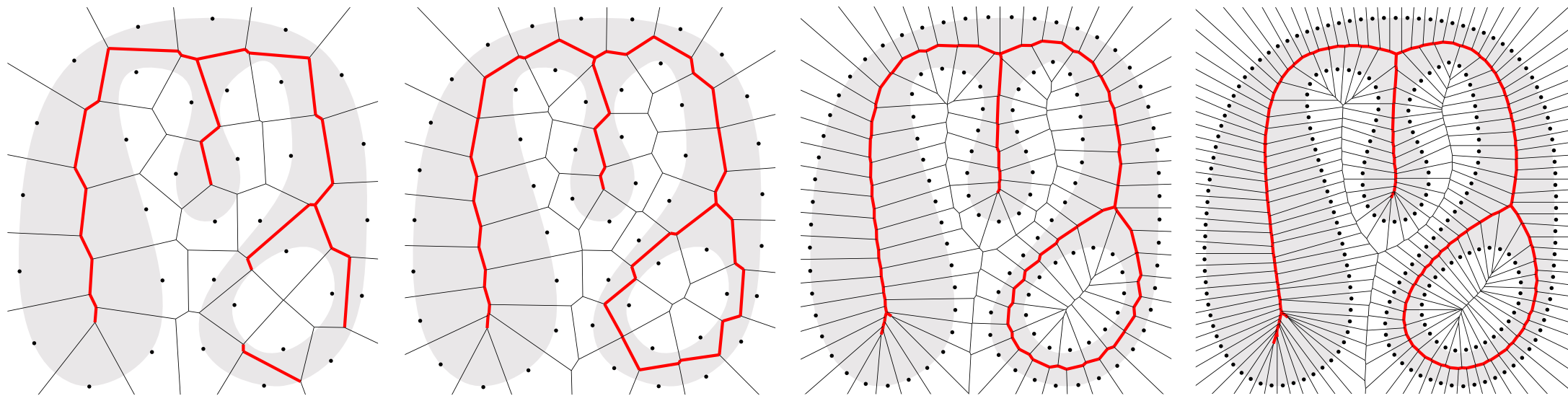}
\caption {The Voronoi diagram of a sampled boundary with increasing uniform sampling density. The Voronoi vertices and edges completely enclosed within the boundary approximate the medial axis. As the density increases, the approximation quality improves. Notice how a minimum sampling distance is necessary to obtain skeletons that are topologically equivalent to the shape.}
\label{fig:medial:voronoi:twod}
\end{figure}

%% file: figures/medial_voronoi_threed.tex
\begin{figure}[th]
\centering
\includegraphics[width=\linewidth]{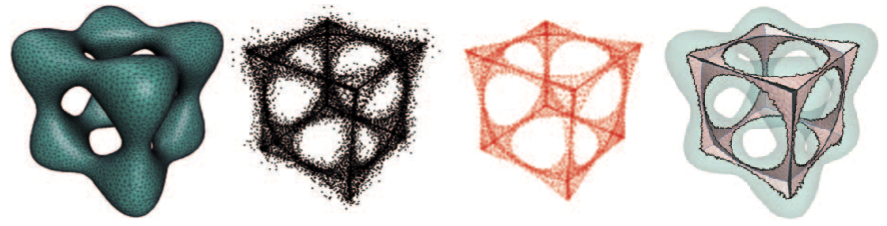}
\caption {\textbf{(a)} A solid object in $R^3$ represented as a triangulation of its surface. \textbf{(b)} The internal subset of Voronoi vertices. \textbf{(c)} The internal subset of Voronoi poles. \textbf{(d)} Given a uniform surface sampling, the $\lambda$-medial axis, a subset of the Voronoi facets, approximate the medial surfaces.}
\label{fig:medial:voronoi:threed}
\end{figure}

%% file: figures/medial_filtering.tex
\begin{figure}[th]
\centering
\includegraphics[width=\linewidth]{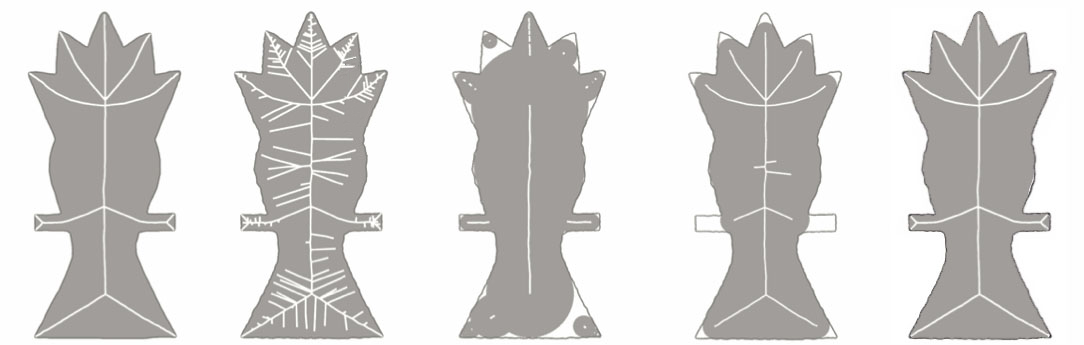}
\caption {\textbf{(a)} A solid planar object and its medial axis. \textbf{(b)} Noise is added to the object boundary resulting in many spurious branches of the medial axis. \textbf{(c)} Filtering medial loci by object angle ($\lambda$-medial axis \protect\cite{chazal_gm05}) captures features across different scales but results in large topological changes. \textbf{(d)} Filtering medial loci by the distance of the corresponding surface points ($\gamma$-medial axis \protect\cite{amenta_sma01}) retains topology but removes small scale features before getting rid of all the spurious branches. \textbf{(e)} The global approach of the scale axis transform is able to remove the noise while retaining small scale features.}
\label{fig:medial:filtering}
\end{figure}

%% file: figures/medial_applications_mix.tex
\begin{figure}[th]
\centering
\includegraphics[width=\linewidth]{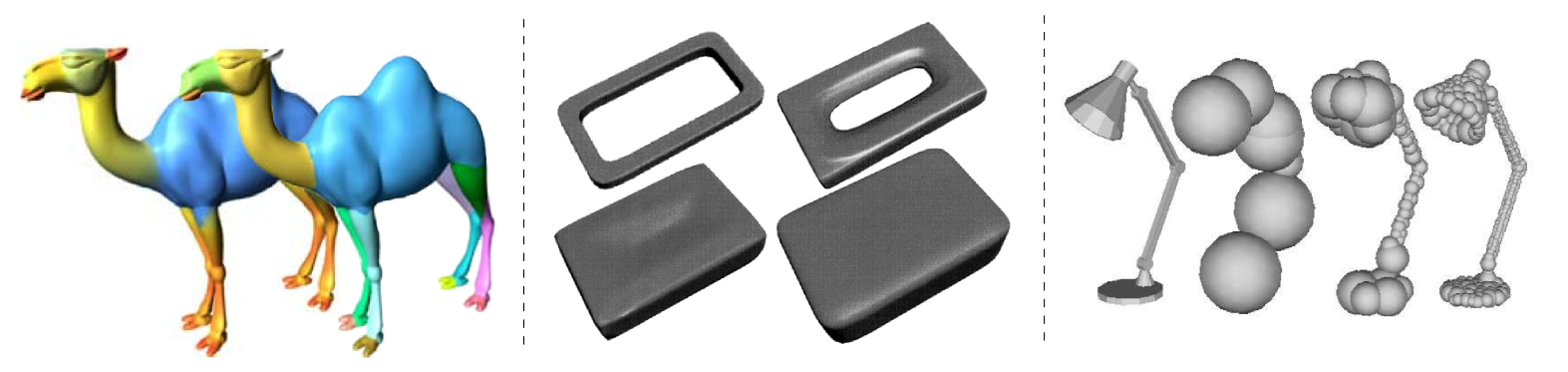}

\caption {\textbf{(a)} The approximation (left) of the local radius in \protect\cite{shapira_vc08} and its application to segmentation (right). \textbf{(b)} Two shapes (on the main diagonal) are blended into each other by exploiting the volumetric structure of the medial axis in \protect\cite{blanding_cg00}. \textbf{(c)} An object represented as a triangulated surface is decomposed by \protect\cite{hubbard_tog96} in a set of spheres to accelerate collision detection.}

\end{figure}

%% file: figures/medial_applications_surfrecon.tex
\begin{figure}[th]
\centering
\includegraphics[width=\linewidth]{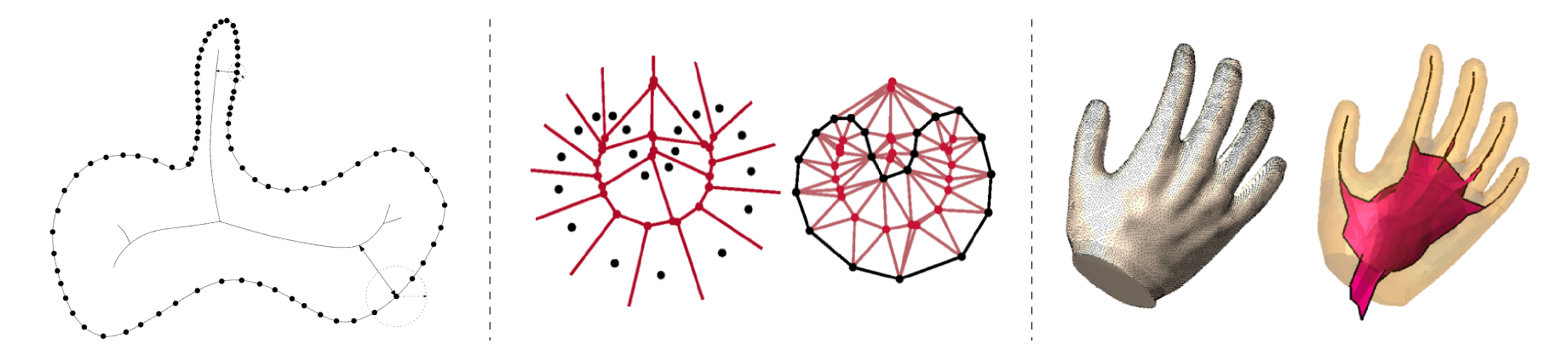}
\caption {\textbf{(a)} An example of the multi-scale $\epsilon$-sampling; note how higher sampling density is required in thin regions, measured by the local feature size. \textbf{(b)} The medial axis is used to discard Delaunay edges connecting opposite sides of the shape's boundary in \protect\cite{amenta_scg98}. \textbf{(c)} An acquired point cloud (left) is reconstructed into a surface and its medial axis (right) by \protect\cite{amenta_sma01}.}
\end{figure}

%% file: figures/medial_applications_shapemanip.tex
\begin{figure}[th]
\centering
\includegraphics[width=\linewidth]{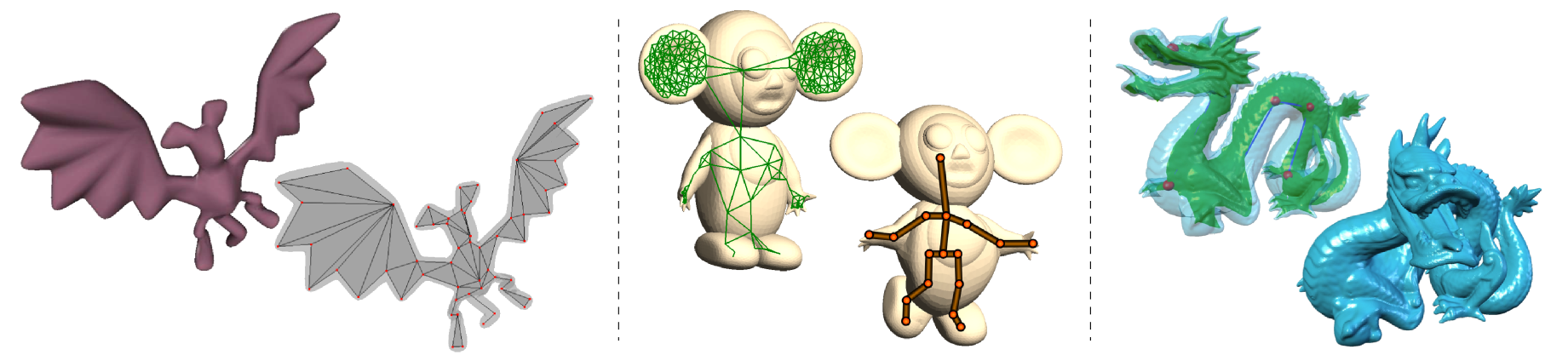}
\caption{\textbf{(a)} An example of a model produced by the medial axis based framework of \protect\cite{angelidis_sma02}. \textbf{(b)} A graph approximating the medial axis is used as an embedding domain for a template skeleton, allowing to re-use animation rigs and produce new poses \protect\cite{baran_sig07}. \textbf{(c)} An example of a deformation achieved by the volume preserving framework of \protect\cite{yoshizawa_eg07}.}
\end{figure}

%% file: skeleton.tex
\section{Curve Skeletons}
\label{sec:skeletons}
\input{figures/skeletons.tex}
\emph{Curve skeletons} were introduced as a way of providing a \emph{curvilinear} (a one dimensional construct) representation of three dimensional shapes. As I have thoroughly discussed in previous sections, this necessity came from the fact that classical medial axis generally produces two-manifold elements when applied to three dimensional shapes. It is important to note that while medial skeletons are precisely and uniquely defined, the same doesn't apply to curve skeletons. There is no universally accepted rigorous definition of what a curve skeleton is. A notable exception is the \emph{medial geodesic skeleton} recently defined by \cite{dey_sgp06}. However, since it is computed as a subset of the medial axis, it also inherits all of its stability issues and has consequently received little attention in applicative scenarios. Without restricting ourselves to this formal definition, we can compute curve skeletons in multiple ways; these different methods will be the subject of \Section{skeletons:computation}. The lack of a common formal background makes the task of comparing these different algorithms quite difficult. An attempt to compare these techniques was performed by \cite{cornea_tvcg07}, where the authors defined a set of \emph{desired} curve skeleton properties and then discussed several algorithms accordingly. In \Section{skeletons:properties} we will briefly discuss these properties, which will find a direct use in creation of curve skeleton extraction (\Section{skeletons:computation}) and their applications (\Section{skeletons:applications}). Note that, similarly to \Section{medial}, I will focus my attention to methods which are suitable for surface-based object representation of 3D geometry. 

\subsection{Properties}
\label{sec:skeletons:properties}
The only strictly necessary property that a \emph{curve} skeleton should have is that it needs to be \emph{thin} -- a one-dimensional entity embedded in the three dimensional domain. As in medial skeletons, it is also often required that this network of curves is \emph{homotopic} to the shape -- that they retain the ability to accurately encode the topological structure of the object they represent. It is also important that curve-skeletons offer \emph{part-awareness}: its branches should provide a one-to-one association with \QUOTE{important} geometric features on the shape. As the graph connecting branches establishes a relationship network, by computing skeletons at different resolutions we obtain a way to represent shape in a \emph{hierarchical} fashion. Additionally, we would like their computation to be \emph{robust}, in the sense that the presence of surface noise should not result in the generation of spurious branches. The \emph{smoothness} of curve skeletons is also important, not only for its visual appeal, but also because it improves compactness, as smooth structures are generally representable with a smaller set of geometric elements. 

Furthermore, as curve skeletons are often employed as representations of solid objects, it is often desirable for its curves to be \emph{contained} within the object. In some scenarios, an ever more restrictive requirement on its positioning is necessary: the curves are required to be \emph{centred} within the object, in a similar way in which MAT skeletons are medial within the shape. The positioning of the medial axis is also critical in situations that require \emph{reliability}: where we ask that every point on the surface is visible from at least one curve skeleton location. It is important to note that, from a geometrical point of view, curve skeletons do not truly provide an alternative shape representation, as we cannot perfectly reconstruct the original shape given a curve skeleton. Curve skeletons are instead to be interpreted as \textit{shape descriptors}: a compact way of capturing essential geometrical and topological information. In order to be able to transfer information from the surface to the curve skeleton it is necessary that proper \emph{correspondences} exist between points on the surface and points on the skeleton. For example, these are essential for the generation of an approximated \emph{reconstruction} of the object. This can be achieved by superimposing a set of simple primitives (e.g. spheres, ellipsoids or by sweeping planar curves) along the curve skeleton. The parameters of these primitives can be inferred by analyzing the corresponding surface geometry. 

It is important to note that not all properties are essential in every application scenario. Furthermore, strictly requiring a skeleton to offer one of these properties might make the satisfaction of others unfeasible. As an example, it is hard to obtain centred (in a medial sense) yet robust skeletons, because the medial axis is sensitive to small surface perturbations. Similarly, we could achieve better reconstruction by introducing extra branches in the representation, albeit at the cost of losing part-awareness.

%
%
\subsection{Curve skeleton computation or \QUOTE{skeletonization}} 
\label{sec:skeletons:computation}
In the literature, curve skeletons are computed or \QUOTE{extracted} following few alternative philosophies. Some algorithms, discussed in \Section{skeletons:computation:medial}, focus on reducing the medial skeleton surfaces down to one dimensional structures. In \Section{skeletons:computation:potential} I will discuss those that attempt to modify the notion of distance in order to create curves, as opposed to surfaces, which lie medially within the shape. In \Section{skeletons:computation:topological} I will move my focus to solutions based on topological analysis, where the core aim is to generate homotopy preserving skeletal representations. Other very successful methods, described in \Section{skeletons:computation:zerovolume}, focus on the fact that we are indeed looking for a curve. I will conclude with \Section{skeletons:computation:grouping}, where I will describe very-application oriented techniques, that construct curve skeletons by grouping surface elements according to particular properties, and then define the skeleton as the graph describing how these components are glued together to form the whole.

\subsubsection{Skeletonization by medial axis processing}
\label{sec:skeletons:computation:medial}
\input{figures/skeletons_computation_dey.tex}
The medial axis of a solid object conveys much of the essential structure of the shape. However, the medial skeleton is geometrically complex as it is composed of a set of surfaces. A simple approach to compute a curvilinear representation of the shape is to prune the medial axis surfaces down to a linear structure. Many heuristics can be used to prune the medial sheets to a line-form; for example, the authors of \cite{lee_gmip94} use a thinning method similar to the one described in \Section{medial:computation:thinning}. A very interesting solution along these lines is the one proposed by \cite{dey_sgp06}, where the authors proposed to extract the skeleton as the shock graph of the \emph{\textbf{M}edial \textbf{G}eodesic \textbf{F}unction}, a scalar function defined on the medial axis surfaces. Given a point $m \in \M$ and its two corresponding surface points $s_1,s_2$, this function is defined as $\text{MGF}(m) = d_\surface(s_1,s_2)$, where $d_\surface(s_1,s_2)$ is the geodesic distance on $\surface$ between the two points. The theoretical argument on the existence of this shock graph provided by \cite{dey_sgp06} make \emph{medial geodesic skeletons} the only mathematically precise version of curve skeletons. Nevertheless, curve skeletons extracted in this fashion suffer severe limitations. Its computational complexity is large, as it necessitates geodesic distances between \emph{every} pair of surface vertices. Furthermore, the sensitivity of medial axis to surface noise, discussed in \Section{medial:properties}, result in the extraction of noisy skeletons like the illustrated in \Figure{skeletons:computation:dey}-d.

\subsubsection{Skeletonization by generalized field analysis}
\label{sec:skeletons:computation:potential}
There exist alternatives to processes that erode the medial axis in order to generate curve skeletons that are well-centred within the shape. These are based on more complex (in particular less localized) ways of measuring the distance from the boundary \cite{hassouna_pami08}. The value of the field at a position in space is computed as the average of potential fields of many boundary samples, and the skeleton extraction is performed by tracing curves seeded at critical points along directions of high divergence. Because of the inherent averaging, these skeletons are not only medial but often very smooth. A major problem with these methods is that they are not particularly suitable for surface based geometry: the field needs to be represented on a discretized volume and, most importantly, they tend to suffer from numerical instabilities. For these reasons they will not be further discussed and I simply refer the reader to \cite{cornea_tvcg07,hassouna_pami08} for more details.

%
\subsubsection{Skeletonization by topological analysis}
\input{figures/skeletons_fieldanalysis.tex}
\label{sec:skeletons:computation:topological}
There are situations in which the core information about an object can be captured by the topological structure of its skeleton (i.e. loops and joints) rather than by its geometric attributes (i.e. curve locations and local radius). In \emph{Morse theory} \cite{shinagawa_cga91}, a real $C^1$ function $\surfacescalar$ defined on the surface $\surface$ of a shape is able to capture its essential topological information: the critical points of $\surfacescalar$ encode the topological changes in the shape. More intuitively, the topological information can be extracted by looking at the iso-contours of $\surfacescalar$ and their distribution on $\surface$ \cite{biasotti_icdg00,natali_gm11} as illustrated in \Figure{skeletons:fieldanalysis}.

In certain situations, like \Figure{skeletons:fieldanalysis}-b the simple $y$ coordinate of boundary points is sufficient. However, there are many ways of creating $\surfacescalar$ and this choice is made according to the type of information we want to extract from the shape. While for a broad comparative analysis we refer the reader to \cite{biasotti_smi03}, for skeleton extraction, a simple solution was introduced by \cite{lazarus_sma99}. Their idea is illustrated in \Fig{skeletons:fieldanalysis}-a, where $\surfacescalar$ is computed as the geodesic distance (or one of its approximations) from a selected source point $p$. Even though the seed point can be selected by heuristics, it is important to note that a single seed generally results in the selection of a preferred slicing direction. An example of the consequences of this choice is illustrated in \Fig{skeletons:fieldanalysis}-a, where we can easily discern a loss of structural detail in the skeleton. To solve these issues, \cite{mortara_algo03} proposed the use of a set of seed points on protrusions, which can be identified by searching for locations of high large gaussian curvature, while other methods like \cite{hilaga_sig01} compute the integral of the geodesic distance to all other surface points (See \Figure{skeletons:fieldanalysis}-b,c).The limitation of all these methods is the necessity to pick an heuristic, as, according to the choice made, unforeseen topological loops can appear in locations where the wavefronts collide. 

It is important to note that this class of algorithms typically offers limited geometric interpretation. The spatial location of a skeletal vertex (i.e. its embedding) is generally computed \emph{a posteriori} -- as the centroid of its corresponding surface points. Consequently, these methods generally offer lower quality geometric skeletons, and are more focused toward providing a correct topological representation. It is very important to note that in applications where topology is the core shape feature, very simple (and efficient to compute) choices for $f$ can be made, like the $y$ coordinate of \Figure{skeletons:fieldanalysis}-b, and these are ensured to produce homotopy preserving curve-skeletons.

\subsubsection{Skeletonization by contraction}
\label{sec:skeletons:computation:zerovolume}
The fundamental requirement of curve skeletons to be a line-like representation can be used for their computation. A line representation can indeed be thought of as a \emph{zero-volume solid}. Thus, in order to be able to extract a skeleton, we can let the shape evolve in such a way that, at each iteration, its total volume is strictly decreased. In order to avoid loss of detail, this volume reduction, commonly referred to as \emph{contraction}, needs to be performed in a controlled fashion. The method used to control contraction is the fundamental ingredient of the algorithms described in this section.

\input{figures/skeletons_computation_wang.tex}
\paragraph{Volumetric contraction}
In \cite{wang_tvcg08}, the authors compute the curve skeleton of a volumetric model by applying a contraction procedure. If we think of the shape's volume as a set of voxel centres connected by edges, it is intuitive to think that we could reduce the shape's volume by uniformly scaling down the edge's lengths. Clearly this isotropic scaling would quickly and meaninglessly shrink the shape down to a zero-volume point as illustrated in \Figure{skeletons:computation:wang}-b. In order to prevent this from happening,  the authors solve a constrained problem where the locations of voxels originally positioned nearby the surface are required not to move too far from their initial location. A fundamental problem in applying this technique to surface models is that it requires a volumetric contraction. Nevertheless the contribution of \cite{wang_tvcg08} is important as it built the groundwork for the very successful surface based contraction method.

\input{figures/skeletons_computation_au.tex}
\paragraph{Surface contraction} In \cite{au_sig08} the authors address the problem of volumetric contraction directly on a meshed surface representation. In the proposed solution they iteratively reduce the volume of the shape (i.e. contraction) by solving a constrained \emph{mean curvature flow} (MCF). Motion by mean curvature is known to result in a volume loss at each iterations \cite{ruuth_jsc03}. While performing MCF on the surface doesn't cause a contraction as isotropic as \cite{wang_tvcg08}, surface attraction constrains need to be inserted in the system to make sure that branches do not undergo excessive shrinkage, thus, resulting in the loss of important surface features. 
The foundations for the success of this technique is that both the volume contraction motion and the feature preserving constraints can efficiently be encoded by a system of linear equations for which efficient solvers are available. Furthermore, as far as a proper Laplacian operator is used, the contraction procedure is not only insensitive to badly discretized surfaces (e.g. low quality triangulation), but since MCF is essentially a smoothing operation, it is also extremely robust to noise.


\paragraph{Topological contraction} While the methods above tackle the problem of contraction in a geometrical sense, it is interesting to know that it's possible to reduce the overall shape volume by only using \emph{topological} operations. In \cite{li_sithreedg01}, the authors propose to use \emph{iterative edge collapse} to reduce a meshed surface to a set of simple segments, which have no incident faces (i.e. wingless). As the resulting set is a simple network of edges, this structure is ensured to be one dimensional. Furthermore, by choosing to collapse the shortest edge first (resulting in an overall $O(n \log n)$ complexity), the authors experimentally demonstrate how this solution results in approximatively-centred skeletons. While no strong theoretical foundations exist for topological contraction, this algorithm lies at the core of many skeletonization algorithms, as it is often employed as a post-processing step: to reduce a quasi-thin structure to a set of truly one dimensional primitives \cite{wang_tvcg08,au_sig08}. It is important to note that in situations in which the quasi-thin structure is already centred within the shape, the topological contraction does not alter the embedding, thus, preserving centeredness.

\subsubsection{Skeletonization by property grouping}
\label{sec:skeletons:computation:grouping}
\input{figures/skeletons_computation_grouping.tex}
If we are given the curve skeleton of a shape, we are provided with a way of decomposing it in fundamental \emph{parts} and their \emph{interconnections}. Note that this relation can be considered in the other direction as well. If our application domain allows us to precisely identify what constitutes a part, then we can construct curve skeletons as the network of curves describing their spatial configuration. There are many ways to define appropriate grouping criteria for skeleton generation and, in this section, I will describe a few important examples.

\paragraph{Rotational symmetry} In \Section{intro} I discussed how one way of interpreting a skeletal axis is by considering it as an \emph{axis of symmetry}. For two dimensional shapes, it was \emph{reflectional symmetry} that allowed us to define medial skeletons -- by grouping pairs of locally symmetric points \cite{zhu_pami99}. In three dimensions a different symmetry relationship can be exploited: \emph{rotational symmetry}. Nevertheless, like in \Section{intro}, the extraction of global rotational symmetry \cite{sun_pami97} would only allow us to define a single global axis. Consequently, the generation of curvilinear representations can be achieved by considering symmetry at a local scale \cite{li_sithreedg01}. The identification of local rotational symmetry (see \Figure{skeletons:computation:grouping}-b) lies at the core of a class of algorithms that extract skeletons by exploiting this geometric observation \cite{chuang_cng04,tagliasacchi_sgp11}.

\paragraph{Segmentation}
Aside from symmetry, there are many ways to partition an object into geometrically meaningful components \cite{shamir_cgf08}. Segmentation schemes which are meaningful in skeletonization usually perform the shape decomposition by following some \emph{convexity} criterion -- by cutting the surface along concave regions. Decomposing a shape into convex components creates a segmentation that is meaningful for skeletonization, as the \emph{principal axis} of a convex component (in a PCA sense) is a linear element that is centred and reliable \cite{lien_spm06} (also see \Figure{skeletons:computation:grouping}-a). Another way of interrelating segmentation to skeletonization is found by considering the animation of articulated objects. For example, the concavity formed by a bent elbow gives us strong cues as to where the articulation takes place. The identification of these concavities is used to decompose the shape into components and allows the generation of hierarchical animation skeletons \cite{katz_sig03}.

\paragraph{Articulation analysis} The skeleton computation algorithms considered up to now mostly addressed static scenarios, where the shape doesn't change in time. However dynamic data can also be exploited in the skeleton extraction process. In \emph{articulated shape animation}, this dynamic data consists of several frames of an object in different poses. These frames can be analyzed to discover the underlying structure. The data contains large connected areas undergoing a similar quasi-rigid transformation. Following this intuition, \cite{anguelov_uai04,schaefer_sgp07,deaguiar_eg08} construct curve skeletons by first clustering parts of the shape undergoing similar transformations, associating a bone to each of them and connecting them together to create an animation rig resembling the ones used in animation (see \Section{skeletons:applications:animation}).

%
%
\subsection{Applications}
\label{sec:skeletons:applications}
Because of their simple yet informative structure, curve skeletons have found uses in many areas including scientific visualization, computer animation, shape analysis, and computer assisted medical diagnosis, etc. The coverage provided by this section is intentionally not exhaustive and it will be strongly biased toward applications in computer graphics and geometry processing.

\subsubsection{Computer animation}
\label{sec:skeletons:applications:animation}
\input{figures/skeletons_applications_animation.tex}

In computer animation, posing digital characters is an essential task and there are multiple ways to achieve it. For example, we can directly manipulate the surface \cite{lipman_tog05} or use control cages \cite{lipman_sig08}. Arguably the most commonly used method is \textbf{S}keletal \textbf{S}ubspace \textbf{D}eformation \cite{lewis_sig00}, where curve skeletons mimic the way real-world bones create deformations in our own bodies. Not only is skeletal deformation found in most 3D modeling packages but its efficient computation is so crucial that it is commonly hardware-accelerated on GPUs. In SSD, the animator specifies a \emph{kinematic} skeleton for a rest-pose of the shape -- like the T-pose of \Figure{skeletons:applications:animation}-(a,left). To obtain a desired pose, a set of transformations (typically rigid) are applied in hierarchical fashion to the rest-pose skeleton. Finally, the posed shape is obtained by transferring the bone transformations onto surface vertices as shown in \Figure{skeletons:applications:animation}-(a,right); this is achieved by performing a weighted combination of transformations. These weights are typically called  \emph{skinning weights}: a scalar field that encodes how much a surface region should be influenced by a particular bone (See \Figure{skeletons:applications:animation}-b). 

As the generation and skinning of kinematic skeletons is a time consuming task, a number of algorithms has addressed the problem of automating the process. For example, in \cite{wade_pca00,dellas_smi07} the authors considered the problem of fitting templates of skeletons in digital human models. In \cite{baran_sig07}, the authors generalized the approach, not only considering 
fitting of more complex skeletons (i.e. not human), but also in the more general sense of transferring animations between different shapes. If a sequence of pre-animated meshes is given, a skeleton can be extracted with the techniques discussed in \Section{skeletons:computation:grouping} and their skinning weights can be computed in a completely automated fashion \cite{schaefer_sgp07} as illustrated in \Figure{skeletons:applications:animation}-c. This possibility of converting animations to the SSD framework has been demonstrated to be very effective for animation compression \cite{deaguiar_eg08}. In addition, note that the generation of kinematic skeletons and weights is not restricted to surface models, but elegantly extends to volumetric data \cite{gagvani_gm01,theobalt_vrst04}.

The applications of curve skeletons in animation are not limited to generating shape poses and parameters. For example, in \cite{lazarus_jvca97}, the authors employ skeletons as a parameterization domain suitable for \emph{shape blending} of simple shapes. Furthermore, the performance of collision detection algorithms based on bounding volume hierarchies can be enhanced by decomposing shapes using skeletons \cite{li_sithreedg01}.

\subsubsection{Geometry processing}
\input{figures/skeletons_applications_geoprocessing.tex}
Curve skeletons have broad applications in the modeling and analysis of geometry. Their branches are part-aware and associated with important shape features, allowing the development of efficient segmentation algorithms for simple human-like objects \cite{xiao_3ddim03,werghi_ieeesmc06}, as well as for shapes with more complex structure \cite{reniers_smi07,tierny_smi07,au_sig08}\ignore{\cite{berretti_ivc09}}. Their ability to compactly represent topology has been exploited to automatically correct topological problems in isosurfaces of implicit functions \cite{wood_tog04} and low quality surface models \cite{zhou_tvcg07}. Its approximate reconstruction properties have been exploited for the modeling of organic shapes \cite{angelidis_smi02,zhongping_pg10} as well as for the correction of geometry in acquired data \cite{tagliasacchi_sig09,cao_smi10,zheng_eg10}. Curve skeletons are also very useful to address the correspondence problem particularly for datasets consisting of shapes characterized by many protrusions. Their compactness allows to greatly reduce the input size of what is essentially a combinatorial problem.

\paragraph{Shape correspondence} Shape correspondence addresses the problem of identifying meaningful relationships between elements of two shapes. A large variety of methods solve the problem by directly analyzing the shape geometry \cite{tangelder_mta08,vankaick_cgf10}. On the other hand, curve skeletons compactly represent a shape as the \emph{graph} of its components, their relationship and, in some scenarios, a rough description of their geometry \cite{biasotti_dgci00}. Consequently, an alternative approach to establish a correspondence between shapes can be obtained by comparing their graphs \cite{hilaga_sig01}. This is advantageous because the size of graph is usually much smaller than the \emph{size} of the surface (i.e. number of triangles) or volumetric representations (i.e. number of voxels), allowing more computationally-expensive algorithms to run in reasonable time. In addition, when shapes exhibit large variation in \emph{poses} and \emph{surface details}, a graph representation better captures the shape's overall structure than local geometric measures (e.g. curvature).

A possible way to compare two graphs is to identify the \emph{maximal common subgraph}, but this problem is NP-hard. Alternative methods employ \emph{graph edit distances} -- by measuring the cost of transforming one graph into the other by simple editing operations. Although this is known to be an NP-hard problem \cite{zhang_ijfcs96}, there exist heuristic solutions (e.g. the graph often needs to be converted into an acyclic one) based on graph editing operations both in 2D \cite{sebastian_pami04} and 3D \cite{biasotti_cad06}. A different way to approach the problem is by noting that the spectral decomposition of the graph's adjacency matrix contains much information about the topology of a graph \cite{siddiqi_ijcv99,sundar_smi03}. As the eigen-decomposition typically captures \emph{global} topological information about the graph, algorithms in this class typically proceed in a coarse to fine manner. For this reason the skeletal graph needs to be not only cycle free but also capable of expressing the natural hierarchical structure of the shape. Other more direct solutions compute correspondences by \emph{directly} comparing sets of skeletal nodes in either a hierarchical way by means of a set of heuristics \cite{hilaga_sig01}, or by a carefully designed voting scheme \cite{au_eg10}.

\subsubsection{Medical shape analysis}
\input{figures/skeletons_applications_medical.tex}
In the medical domain, curve skeletons are well suited to describe anatomical structures that have characteristic tubular shapes like vessels, nerves and elongated muscles \cite{nystrom_iciap01,fridman_mia04}. The compactness of the representation allows to efficiently perform computational tasks like registration of partially overlapped vessel images \cite{aylward_ijcv03} and the flattening of their three dimensional structure to a plane \cite{kanitsar_vis03}. Furthermore, the encoded local volumetric information can help in the detection of abnormalities in vascular structures, like stenosis \cite{sorantin_tmi02} and aneurisms \cite{straka_vis04}. Skeletons satisfying centeredness and reliability are also particularly important in medical applications. Indeed, in virtual navigation \cite{wan_vis01}, it is essential that the physician is able to examine the full internal surface of an organ \cite{he_tvcg01}. In this context, appropriately extracted curve skeletons can be used to generate camera fly-through paths for the inspection of the intestine \cite{hong_sig97,wan_tmi02}, lungs \cite{perchet_spie04} and blood vessels \cite{bartz_vis99}. In addition, when the structure is so tangled that even a virtual navigation would be uninformative, the skeleton can be used to straighten the organ to a linear structure which can be more easily analyzed \cite{silver_mmvr02}.

%% file: figures/skeletons.tex
\begin{figure}[th]
\centering
\includegraphics[width=\linewidth]{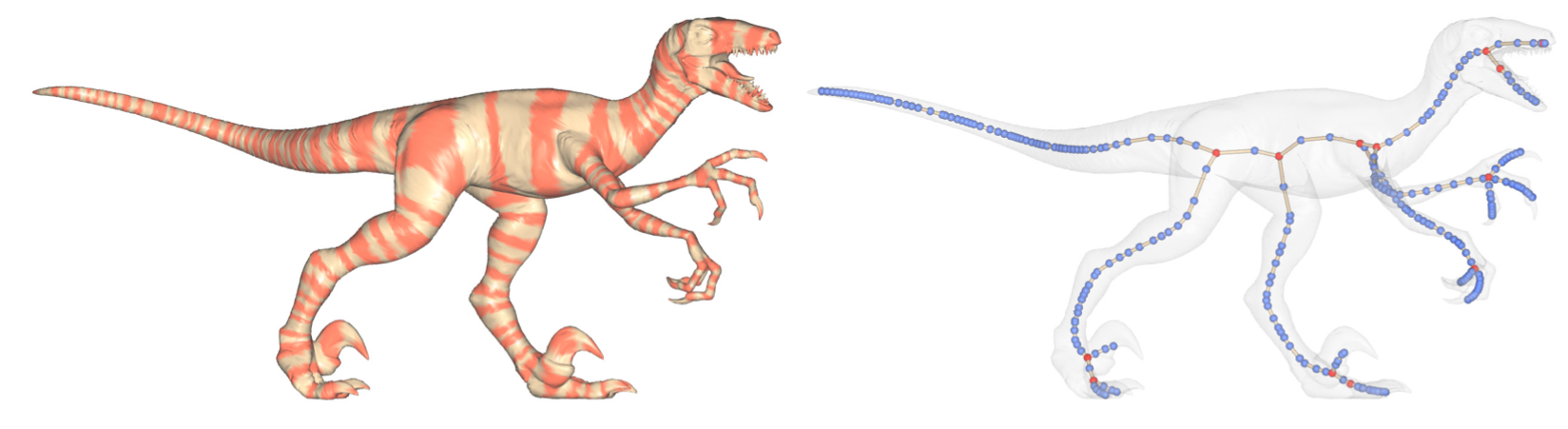}    
\caption {\textbf{(a)} A three dimensional object; the colour coding expresses the correspondence between portions of the surface and vertices on the curve skeleton. \textbf{(b)} An example of a curve skeletal representation extracted by \protect\cite{au_sig08}.}
\label{fig:skeletons}
\end{figure}

%% file: figures/skeletons_computation_dey.tex
\begin{figure}[th]
\centering
\includegraphics[width=\linewidth]{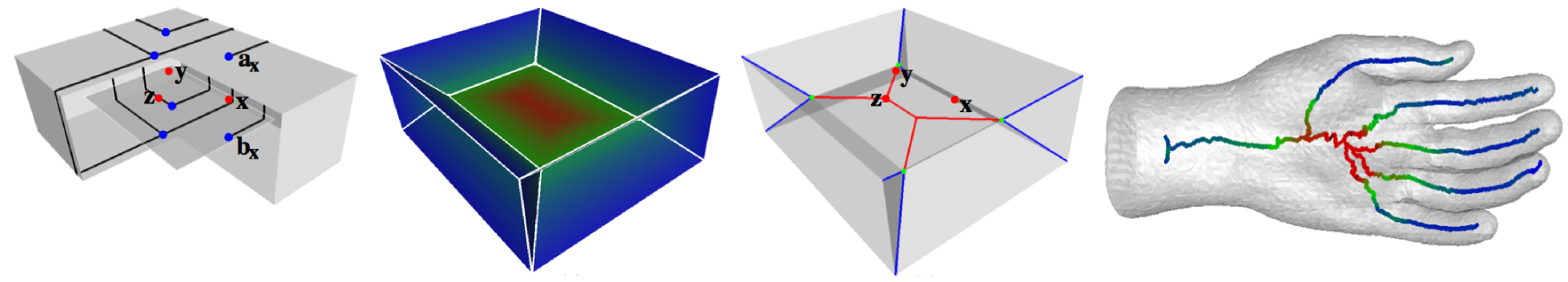}
\caption{An illustration of the fundamentals of the described in \protect\cite{dey_sgp06}. \textbf{(a)} Three points (red) on the medial axis of a cuboid; their corresponding surface points (blue), and the shortest paths on the surface between these points (black). \textbf{(b)} The medial geodesic function on the medial axis of shape. \textbf{(c)} High divergence points of the medial geodesic function define the curve skeleton. Note how only $y$ and $z$ are on the skeleton as they are associated with more than one shortest path. \textbf{(d)} The curve skeleton of a noisy hand model is also noisy as it is defined as a subset of the medial axis.}
\label{fig:skeletons:computation:dey}
\end{figure}

%% file: figures/skeletons_fieldanalysis.tex
\begin{figure}[th]
\centering
\includegraphics[width=\linewidth]{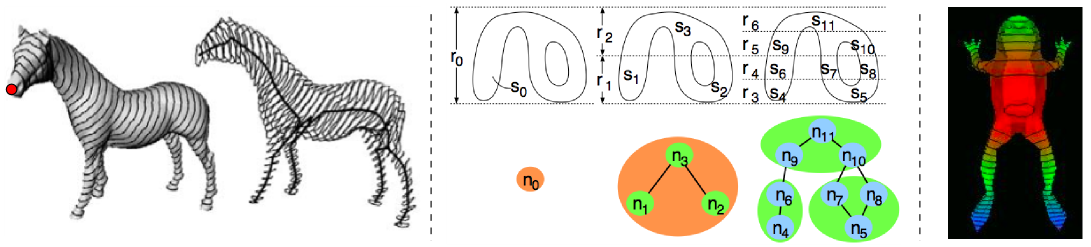}
\caption {\textbf{(a)} A geodesic distance field sourced at the marked point is created onto the surface and its iso-contours are shown. The barycenter of surface portions between two successive iso-contours is used to build a curve skeleton like embedding of the topological graph \protect\cite{lazarus_jvca97}. \textbf{(b)} The $y$ coordinate is used as the field for topological analysis. By recursively refining its iso-contours, a multi-resolution topological graph is created \protect\cite{hilaga_sig01}. \textbf{(c)} The iso-contours of the average geodesic function from \protect\cite{hilaga_sig01} naturally capture the extremities of the articulated shape of a frog.}
\label{fig:skeletons:fieldanalysis}
\end{figure}

%% file: figures/skeletons_computation_wang.tex
\begin{figure}[th]
\centering
\includegraphics[width=.95\linewidth]{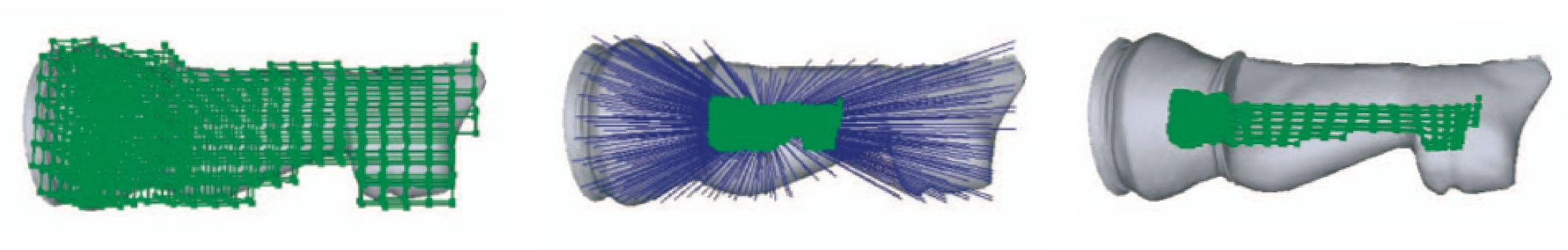}
\caption {An illustration of the volumetric contraction proposed by \protect\cite{wang_tvcg08}. \textbf{(a)} The surface is first converted into a volumetric representation. \textbf{(b)} The edges of the volume are isotropically contracted, but by using the surface attraction constraints (blue) we can avoid the shape from collapsing to a point. \textbf{(c)} By using these attractors the collapsed volume retains the shape features.}
\label{fig:skeletons:computation:wang}
\end{figure}

%% file: figures/skeletons_computation_au.tex
\begin{figure}[th]
\centering
\includegraphics[width=\linewidth]{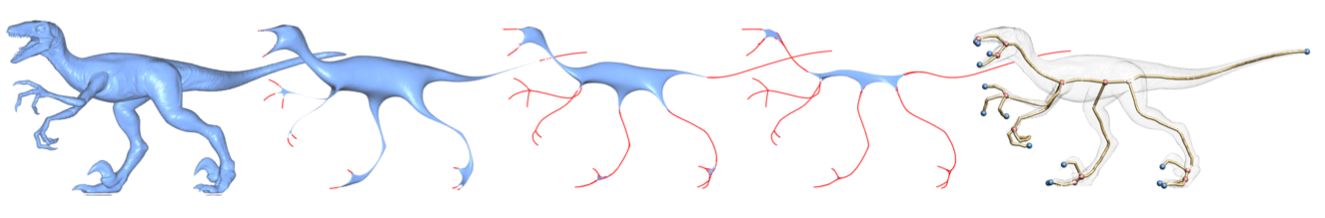}
\caption {An illustration of the surface contraction proposed by \protect\cite{au_sig08}. \textbf{(a,b,c,d)} From left to right the shape is contracted by applying curvature flow until the volume vanishes. To prevent over-contraction surface attraction constraints similar to \protect\cite{wang_tvcg08} are applied. \textbf{(e)} The contracted surface retains the original manifold topology and is collapsed to linear graph  \protect\cite{li_sithreedg01}.}
\end{figure}

%% file: figures/skeletons_computation_grouping.tex
\begin{figure}[th]
\centering
\includegraphics[width=\linewidth]{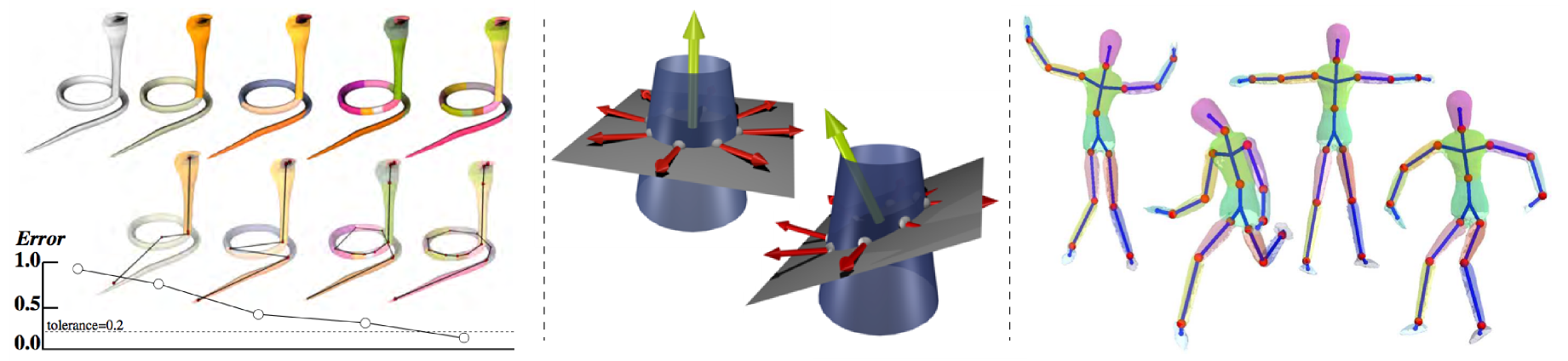}
\caption{\textbf{(a)} A convex decomposition allows to recycle PCA axes as skeletal branches \protect\cite{lien_spm06}; the finer the decomposition, the more accurate the skeletonization. \textbf{(b)} Local rotational symmetry of a group of samples can be measured by observing the arrangement of the surface normals on the gaussian image \protect\cite{tagliasacchi_sgp11}. \textbf{(c)} The articulation of a shape is essentially described by a set of locally rigid transformations \protect\cite{anguelov_uai04}; these transformations can be identified and exploited to create a kinematic skeleton.}  
\label{fig:skeletons:computation:grouping}
\end{figure}

%% file: figures/skeletons_applications_animation.tex
\begin{figure}[th]
\centering
\includegraphics[width=\linewidth]{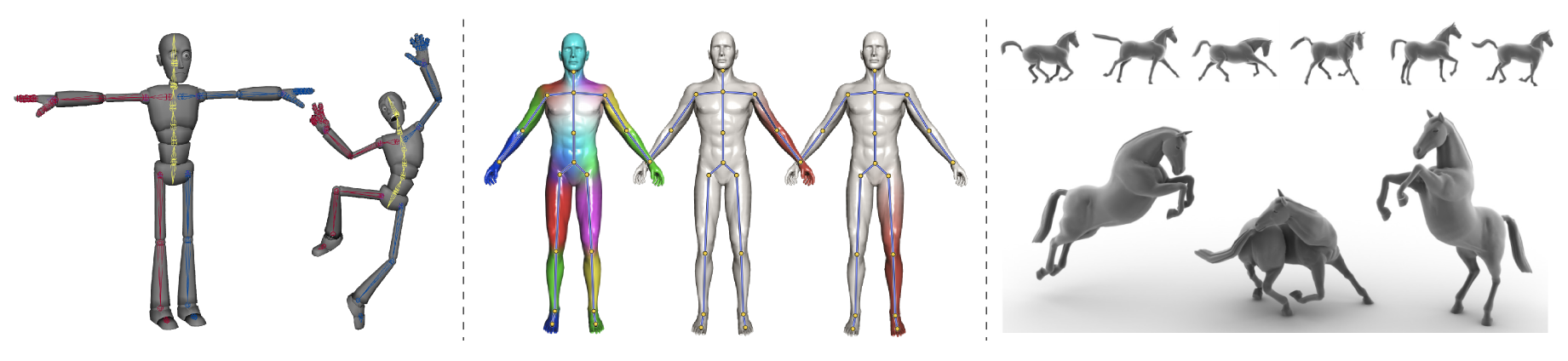}
\caption {\textbf{(a)} An example of a kinematic skeleton (also called animation rig) from MAYA is used to articulate the shape from its rest-pose to a desired one. \textbf{(b)} Each bone (i.e. edge) of the curve skeleton is associated with skinning weights that describe how its movement affects the surface vertices \protect\cite{jacobson_sig11}. \textbf{(c)} Given a set of example poses (top),  it's possible to automatically estimate skinning weights and skeletons \protect\cite{schaefer_sgp07}; (bottom) this information is sufficient to extrapolate new poses.}
\label{fig:skeletons:applications:animation}
\end{figure}

%% file: figures/skeletons_applications_geoprocessing.tex
\begin{figure}[th]
\centering
\includegraphics[width=\linewidth]{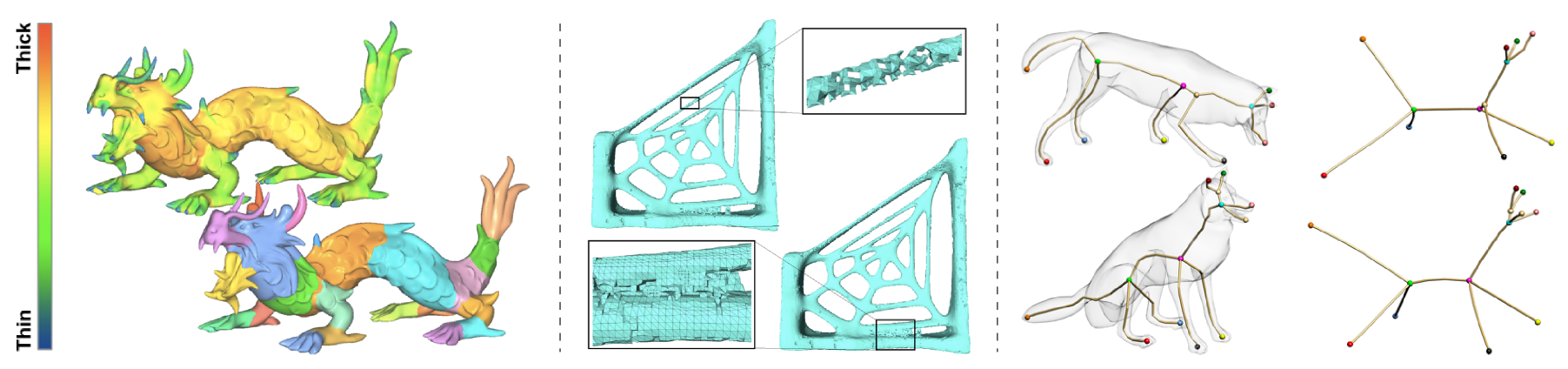}
\caption {\textbf{(a)} The surface to skeleton correspondences allow to define a radius function (top) onto the surface which can be used to segment the shape (bottom) \protect\cite{au_sig08}. \textbf{(b)} The compact topological representation of the curve skeleton is used to drive a topology fixing process \protect\cite{zhou_tvcg07}. \textbf{(c)} Multi-dimensional scaling can be efficiently performed on the skeleton, resulting in stretched out limbs which facilitate the task of left/right disambiguation commonly found in articulated shape correspondence \protect\cite{au_eg10}.}
\end{figure}

%% file: figures/skeletons_applications_medical.tex
\begin{figure}[th]
\centering
\includegraphics[width=\linewidth]{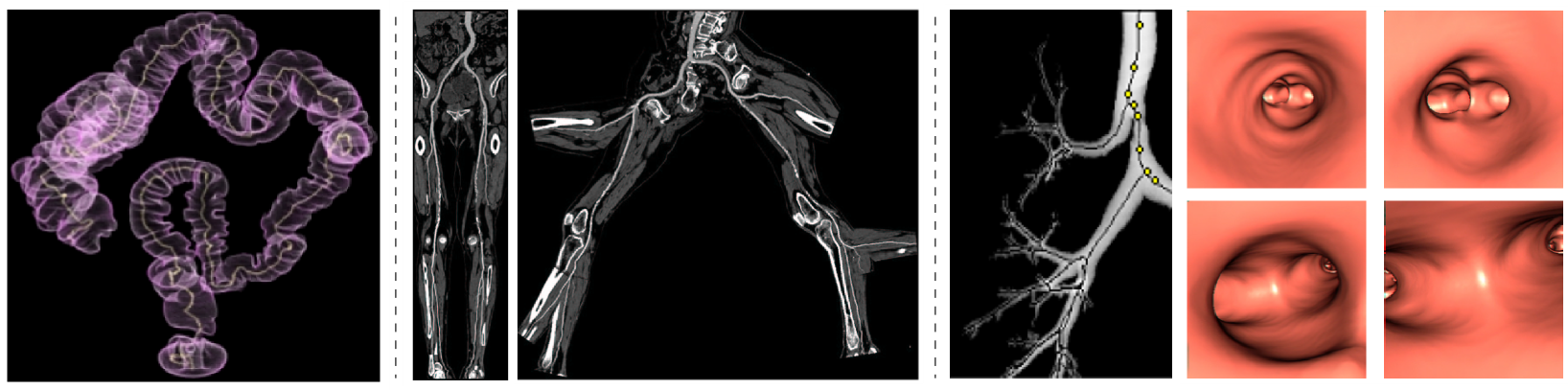}
\caption {Examples of the use of curve skeletons in medical applications. \textbf{(a)} The skeletal centreline of the colon is extracted to allow virtual colonoscopy \protect\cite{wan_vis01} or its unwinding \protect\cite{silver_mmvr02}. \textbf{(b)} The skeleton of the vessel structure is used to perform planar reformation to visualize its untangled structure \protect\cite{kanitsar_vis03}. \textbf{(c)} The centreline automatically generates flight paths for virtual bronchoscopy \protect\cite{perchet_spie04}. }
\end{figure}

%% file: conclusions.tex
\section{Conclusions}
In this report, I presented a detailed overview of skeletal representations as an alternative (or addition) to boundary or volumetric representations for shape processing. 

First, I introduced the idea of \emph{medial skeletons} as a way of capturing redundancy and obtaining local axial representations. I discussed different possible interpretations, several classes of algorithms for its computation, its limitations in terms of stability, along with several ways to tackle this problem. For surface-based representations, I justified how Voronoi-based medial axis extraction methods offer clear advantages with regards to their competitors: not only do they offer theoretical convergence guarantees under appropriate sampling, but efficient and robust libraries for three dimensional voronoi computation are freely available as well. Although, only for two dimensional shapes, medial skeletons result in curvilinear structures, I illustrated how the medial surfaces associated with three dimensional solids contain useful information exploitable in many applications.

The necessity for a simpler curvilinear structure for three dimensional shapes stirred the discussion toward \emph{curve skeletons}. While there is no unique definition for curve skeletons, I concisely described the properties that these curvilinear representations are commonly required to have. I then presented several classes of algorithms for skeletonization of solid shapes, focusing on those algorithms that operate on boundary representations. I highlighted how contraction methods currently represent the state of the art for skeletonization, because they can efficiently extract skeletons from low quality and possibly noisy surface representations without needing any domain-specific information. I concluded my discussion by describing a number of applications of curve skeletons and highlighting their abilities in describing shape from various applicative fields.

%% file: appendix.tex
\newpage
\appendix
\section*{Appendix}

\paragraph*{Maximal inscribed ball}
\label{app:maxball}
Let $\object$ be a closed connected set in $\R^n$. A closed ball $B \subset \R^n$ is called \emph{maximal inscribed ball} in $\object$ if $B \subset \object$ and there does not exist another ball $B' \neq B$ such that $B \subset B' \subset S$.

\vspace{-.15in}
\paragraph*{Voronoi diagram}
\label{app:voronoi}
Given a finite set of points $S$ in $\R^k$, for each point $p$ in $\R^k$ there is at least one point in $S$ closest to $p$. a point $p$ may be equally close to two or more points in $S$. For each point in $S$ its \emph{Voronoi cell} is defined as the subset of $R^k$ of points closest to it than to any other point in $S$. The union of Voronoi cells of all points in $S$ is called the \emph{Voronoi Diagram} of the set $S$. For $n=|S|$ and $k=3$ the complexity is $\Uptheta(n^2)$, but when points are well distributed on a smooth surface the  complexity has been shown to reduce to $O(n \log n)$ \cite{attali_scg03}.

\vspace{-.15in}
\paragraph*{Voronoi pole}
\label{app:voronoipole} 
Given a finite set of points S in $\R^k$ and its Voronoi diagram, each sample point $p$ is associated with a convex Voronoi polyhedron. The vertices on the polyhedron on the two sides of the boundary which are further from $p$ are the \emph{Voronoi poles} of $p$. Note that there are at most $2n$ poles for a sample of size $n$.

\vspace{-.15in}
\paragraph*{Local feature size \ignore{\cite{amenta_cgta00}}}
\label{app:localfeaturesize} 
The local feature size $\text{LFS}(x)$ is the minimum Euclidean distance from any point x to any point of the medial axis.

\vspace{-.15in}
\paragraph*{Delta sampling (i.e. uniform sampling)} 
\label{app:deltasampling}
A sample $P$ of $\surface$ is a $\delta$-sample if no point $p \in \surface$ is further than $\delta$ from $P$.

\vspace{-.15in}
\paragraph*{Epsilon sampling (i.e. adaptive sampling)}
\label{app:epsilonsampling}
A sample $P$ of $\surface$ is an $\epsilon$-sample if no point $p \in \surface$ is farther than $r \cdot LFS(p)$ from $\M$.